\documentclass[12pt]{article}

\usepackage{latexsym,amsmath,amssymb,theorem,epsfig}

\usepackage{cite}

\DeclareFontFamily{OT1}{rsfs10}{}
\DeclareFontShape{OT1}{rsfs10}{m}{n}{ <-> rsfs10 }{}
\DeclareMathAlphabet{\mathscript}{OT1}{rsfs10}{m}{n}

\numberwithin{equation}{section}

\newcommand{\ns}{\normalsize}

\def \rX {{\rm X}} 
\def\a{\alpha}

\def\k{\kappa}
\def\l{\lambda}
\def\m{\mu}
\def\n{\nu}

\def\p{\pi}

\def\s{\sigma}

\def\x{\xi}

\def \S {{\cal S}}

\def\gsim{ \lower .75ex \hbox{$\sim$} \llap{\raise .27ex \hbox{$>$}} }
\def\lsim{ \lower .75ex \hbox{$\sim$} \llap{\raise .27ex \hbox{$<$}} }
\def\be{\begin{equation}}
\def\ee{\end{equation}}
\def\bea{\begin{eqnarray}}
\def\eea{\end{eqnarray}}

\def \rx {{\rm x}}

\def \V  {{\rm V}}
\def \ha {{1 \ov 2}}

\def \sql {{\sqrt{\l}}\ }
\def \del{\partial}
\def \a {\alpha}

\def\ov{\over}
\def \ci {\cite}
\def \n {{\rm n}}
\def \cc {{\hat c}}
\def \hD {{\hat D}}
\def \foot {\footnote}
\def \bi{\bibitem}
\def\la{\label}\def\foot{\footnote}\newcommand{\rf}[1]{(\ref{#1})}

\theoremstyle{plain}

{\theorembodyfont{\rmfamily} }

\def \ws {{\rm  A}}

\def \cD {{\cal D}} 

\def \no {\nonumber}

\begin{document}


\begin{titlepage}

\vspace{-5cm}


\vspace{-5cm}

\title{
  \hfill{\small Imperial-TP-AT-2010-05  }  \\[1em]
   {\LARGE  Semiclassical four-point  functions in  $AdS_5 \times S^5$}
\\[1em] }
\author{
   E.I. Buchbinder  and   A.A. Tseytlin
     \\[0.5em]
   {\ns The Blackett Laboratory, Imperial College, London SW7 2AZ, U.K.}} 

\date{}

\maketitle

\begin{abstract}
We consider a semiclassical (large string tension $\sim \sql$) limit 
of  4-point correlator of two ``heavy''  vertex 
operators with large quantum numbers and two ``light'' operators.
It can be written in a factorized  form as a product of two 3-point 
functions, each  given  by the integrated ``light'' vertex operator  on the classical  string 
solution determined by the ``heavy'' operators. We check consistency of
this  factorization in the case of a correlator  with two dilatons 
as ``light'' operators. We study in detail  the  example
when  all  4 operators  are  chiral primary scalars, two of which carry
large  charge  $J$ of order of string tension.  In the large $J$ limit this
correlator is nearly extremal. Its  semiclassical expression is, indeed, 
found to be consistent with  the general protected form expected for an 
extremal correlator. We demonstrate explicitly that our
semiclassical result matches  the large $J$ limit of the known  
free ${\cal N}=4$ SYM  correlator for 4 chiral primary operators with 
charges $J,-J,2,-2$; we also  compare it  with
an existing supergravity expression.  As an example of a  4-point function 
with  two  non-BPS ``heavy'' operators, we consider  the case  when 
the latter are  representing folded  spinning with 
large $AdS$  spin and two  ``light'' states  being  chiral primary scalars. 

\end{abstract}

\thispagestyle{empty}

\vskip  65pt

\centerline{\it Dedicated to the memory of V.Ya. Fainberg}

\end{titlepage}

\def \adss {$AdS_5 \times S^5\ $}

\def \a {\alpha }


\section{Introduction}

A correlator of operators in \adss string theory   carrying large charges 
of order of string tension ($\sim \sql$)   should be dominated 
 at large  $\l$  by its   semiclassical limit. This  observation was used 
  in the past (see, e.g.,  \ci{pol,gkp2,zar,tsev,tsu,russ})
 and was recently  applied  to computation of 
 2-point \ci{b1,jan,bt1}  and certain  3-point   \ci{z,cost,rt,ry,rut,gro}
 correlators of string vertex  operators.
 
 
 The  main idea \ci{z,cost}  is that a 
 special  subset  of 3-point correlators   containing  
 only two ``heavy'' operators  with   large
 quantum  numbers can be computed 
 using the same  stationary point trajectory that 
 controls the  semiclassical  limit of their 2-point function.
 
In \ci{rt} this   observation was further generalized and  applied to the case when 
 the  ``light'' operator may  be  representing a string mode 
 (i.e. may not be BPS). 
  It was  also  suggested \ci{rt} that the same 
 approach should apply to higher $n$-point  correlation functions
  with 2 ``heavy''  and $n-2$  ``light'' operators: the semiclassical expression 
  for  $n$-point  correlator 
   should be  given by a product of ``light'' vertex 
  operators  computed on  the semiclassical world surface
  determined by the ``heavy'' operator insertions.

Our aim  here  will be to  study in detail  the case 
  of such  4-point  functions, providing evidence of consistency 
  of the
  semiclassical recipe for their computation at strong coupling. 
   Using  an independent argument 
 based on differentiating over string tension, we will show, 
 following \ci{rt}, 
   that the semiclassical  
 expression  for the 
 4-point correlator with  2  integrated dilaton operators 
 can be  represented as a product of two 3-point correlation functions, each 
 with  one dilaton  operator, which   matches   the semiclassical 
 prescription. 
 
 Below we  will consider several  explicit examples.
   In particular, we   will   find  an   explicit form  of   the semiclassical 
   4-point function involving  two ``heavy'' 
 operators corresponding to large-spin folded string in $AdS_5$ 
 and two ``light'' chiral primary scalar operators.
 
 We will also  consider   the case when  the 4-point 
 function  contains  two ``heavy'' and two  ``light'' 
 chiral primary  scalar operators with large charges $\pm J$
 and  fixed charges $\pm j$ respectively. 
   Since  such  correlator is close to be extremal \ci{106}
  for $J \gg j$ one  may expect  that 
 it may   be protected  for large $J$, just like the chiral primary 3-point function 
 is \ci{Lee}. Indeed, we will find  that 
 it exactly matches the $J \gg 1$  limit of the   free
  gauge theory  result \ci{Rayson}
 for the correlator  of 4 chiral primary operators with  charges 
 $J, -J, 2, -2$. 
 
 One may also expect that the semiclassical (large $\l$, large charge)  limit 
 of a  correlation  function of   4  BPS operators should 
 match the large charge limit of the  corresponding supergravity 
 expression computed according to   the standard AdS/CFT rules
 (see, e.g., \cite{Freedman}). Analysing the large $J$ 
 limit  of the  supergravity  expression for the $(J, -J, 2, -2)$
 correlator found in  \cite{Linda1} we will, however, find a
   disagreement 
 (the supergravity correlator grows slower with $J$ than 
 its semiclassical or gauge theory  counterparts). This issue deserves  further
 investigation. 
 
 \
     
The rest of the  paper is organized as follows. In Section 2 we
consider 
 4-point functions at strong coupling in the  case 
when two operators are ``heavy'', i.e. represent semiclassical
states, and show that they can be written  in a factorized 
form as a product of two 3-point functions. 
In Section 3 we give an independent 
proof of that factorization in the case when the ``light'' operators 
are   the dilaton operators, thus  providing a 
 check of the semiclassical prescription. 

In Section 4 we discuss  several  examples of semiclassical 
computation of 
4-point functions.
 We start in section 4.1 with reviewing the general form of the chiral 
 primary vertex operator from \ci{z,ber,Lee} 
 and revisit   the computation  \ci{z,rt}
 of the 3-point correlator  of  chiral primary scalars 
 in the case when two of them carry large charge. 
 Keeping the $AdS_5$ boundary  positions of the operators arbitrary
   helps to clarify, in section 4.2,  the   factorized structure of 
 the semiclassical  4-point function of 2 ``heavy'' and 2 ``light''
 chiral primary operators. 
 Assuming the charge  assignment such that this correlator is an 
 extremal one, we show that this factorized semiclassical structure is in 
perfect  agreement with the   non-renormalization conjecture \ci{106}
 for the extremal correlators.
Finally, in section 4.3  we apply the 
semiclassical  method to compute 
the 4-point function with  2 ``heavy''   large $AdS_5$ spin operators 
and 2  ``light'' chiral primary scalars. 

In section  5 we compare our semiclassical expression for
the  4-point function of  chiral
primary operators  with the  known  free gauge theory ~\cite{Rayson}
and the supergravity  ~\cite{Linda1}
expressions  for
the chiral primary  correlator with charges $J, -J, 2, -2$. 
Taking the  
 large $J$ limit  we find   perfect  agreement with the gauge theory result but an 
apparent disagreement with the supergravity expression  of ~\cite{Linda1}.

Section 6 contains some  remarks  on consistency between the semiclassical 
result  and general  factorization properties  of 4-point functions. 
We also mention some generalizations. 

In Appendix A we present the general form  of the 
chiral primary vertex operator of \ci{z,ber} and explain when 
it can be replaced by its simplified  form used in \ci{rt,rut}.
In Appendix B we consider the large $J$ limit of an $AdS_5$ 
integral entering the supergravity
correlation  function  discussed  in section 5.




\section{Semiclassical  correlation  functions in $AdS_5 \times S^5$ with two
``heavy''  operators}


Our  object of interest in this paper is  4-point  
correlation  function of
string vertex operators dual to 
 gauge invariant 
local operators
with two  operators  carrying  large quantum numbers of order of string tension 
and the remaining $2$   carrying   fixed (much smaller) quantum numbers. 
We will refer to the former two operators as ``heavy'' (or ``semiclassical'') and to the
 latter   as  ``light'' (or ``quantum''). 
 In general, one may consider similar $n$-point correlators with 
   any  number  $n-2=0,1, 2,3, ...$ of such 
 ``light'' operators.
 We will start  with reviewing the  case of the 
two-point ~\cite{b1, jan, bt1} and three-point correlation 
functions~\cite{z, rt}.

The calculation of two-point function  in the leading  semiclassical approximation 
was shown in~\cite{tsev,b1,bt1} to be intrinsically related
to finding an appropriate classical string solution.
Let $ V_{H1}(\xi_1)$ and $ V_{H2}(\xi_2)$ be the 
two ``heavy'' vertex operators inserted at points
$\xi_1$ and $\xi_2$ on the worldsheet (chosen as  a plane 
or a sphere).\footnote{We consider  planar AdS/CFT duality, 
i.e.  only tree-level  string theory (world sheets of higher genera correspond to including 
string or $1/N$ corrections).}
For large string   tension 
($\sql \gg 1$) 
 two-point function in the semiclassical approximation is dominated by the action 
evaluated at  its stationary point
\be
\langle  V_{H1}(\xi_1) V_{H2} (\xi_2)\rangle \sim e^{-I}\,,
\label{1.1}
\ee
where  $I$ is the string action on $AdS_5 \times S^5$ in conformal gauge
($\a=1,2$)
%
\bea
&&I=\int d^2 \xi\  L \ , \ \ \ \ \ \ \ \ \
L=  \frac{\sqrt{\lambda}}{4\pi} (\partial_\a Y_M { \partial_\a} Y^M
+ \partial_\a X_k {\partial_\a} X_k +{\rm fermions})\,,\la{st}\\
&&Y_M Y^M =-Y_0^2 -Y_5^2 +Y_1^2  +Y_2^2  +Y_3^2  +Y_4^2 =-1\,, \nonumber\\  
&&X_k X_k =X_1^2 +\ldots + X_6^2 =1\,.
\label{1.1.1}
\eea
The stationary point solution  solves the 
 string equation with singular sources, i.e. has singularities  
 prescribed by 
$V_{H1}(\xi_1)$ and $V_{H2}(\xi_2)$. 
Using conformal symmetry we can map the $\xi$-plane 
to the Euclidean cylinder parametrized by $(\tau_e, \sigma)$
\be
e^{\tau_e +i \sigma}=\frac{\xi-\xi_2}{\xi-\xi_1}\,.
\label{1.2}
\ee
It was shown on  various examples in flat space and in $AdS_5 \times S^5$ 
in~\cite{tsev,b1,bt1} that under this conformal map the 
singular solution on $\xi$-plane 
transforms into a smooth classical  string
solution on the cylinder that 
 carries the same quantum numbers (energy, spins, etc.)
as the  states  represented by the  vertex operators.

This discussion  can be repeated 
for the physical   integrated  vertex operators  labelled 
by  points $\vec{{\rm x}}_1$, $\vec{{\rm x}}_2$ on the boundary of the 
Poincar\'e patch of  $AdS_5$  
\cite{pol,tsev}
\be
\V_{H}(\vec{{\rm x}})=\int d^2 \xi \  V_{H}(\x ; \vec{{\rm x}}) \ , \ \ \ \ \ \ \ \ \ \ 
V_{H}(\x ; \vec{{\rm x}})\equiv V_H(z (\xi), \vec{x} (\xi)-\vec{{\rm x}}, X_k(\xi))\,,
\label{1.3}
\ee
where $z$ and $\vec{x}=(x_{0e}, x_1, x_2, x_3)$ are the Poincar\'e coordinates 
of $AdS_5$.\footnote{Throughout this  paper $AdS_5$ is always assumed to be Euclidean.}
The semiclassical two-point function 
$\langle V_{H1}(\vec{{\rm x}}_1) V_{H2} (\vec{{\rm x}}_2)\rangle$
is again determined by the value of the classical action on the stationary point solution. 
%
%
 After  we perform the conformal map~\eqref{1.2} we again obtain a smooth 
solution on the cylinder which is just the corresponding spinning string solution
rewritten in Poincar\'e coordinates  which  will satisfy 
the following boundary conditions (see \cite{bt1} for details)\footnote{
In general, the euclidean solution will not be real and  is not required 
to end on points at the boundary. For example, for an operator representing  long folded 
spinning string  the  corresponding solution 
will approach 
 null lines passing through the insertion 
points  \cite{bt1}.  This  subtlety  will not be important for what follows.}
\be
\tau_e \to -\infty  \ \ \Longrightarrow \ \  z\to 0,\  \  \vec{x}\to \vec{{\rm x}}_1\,,
\ \ \ \ \ \ \ \ \ \ \ \ \ \ \tau_e \to +\infty  \ \  \Longrightarrow \ \  z\to 0,\  \  \vec{x}\to \vec{{\rm x}}_2\,.
\label{1.5}
\ee
%
Similar  method  can be  applied  ~\cite{z, rt} 
to the  semiclassical computation  of three-point functions
with  two ``heavy'' and one  ``light'' operators
\bea
&&G_3 (\vec{{\rm x}}_1, \vec{{\rm x}}_2, \vec{{\rm x}}_3)=
\langle \V_{H1}(\vec{{\rm x}}_1) \V_{H2}(\vec{{\rm x}}_2)\V_L(\vec{{\rm x}}_3)\rangle
\nonumber\\
&&=\int {\cal D} {\mathbb X}^{\mathbb M}\ e^{-I} \ \int d^2 \xi_1 d^2 \xi_2d^2 \xi_3\
 V_{H1}(\xi_1; \vec{{\rm x}}_1) 
V_{H2}(\xi_2; \vec{{\rm x}}_2) V_L(\xi_3; \vec{{\rm x}}_3)\,,
\label{1.6}
\eea
where $\int {\cal D} {\mathbb X}^{\mathbb M}$ is the integral over  the fields
$(Y_M, X_k)$  (as well as fermions which we ignore as we consider only leading-order 
semiclassical expansion). 
 In the stationary 
point equations 
 the contribution of the 
 ``light'' operator can be
ignored and then the  solution is the same as 
in the case of two-point function of two ``heavy'' operators, i.e. 
%
%
it is found by extremising  \be 
I- \ln V_{H1}(\xi_1; \vec{{\rm x}}_1) -\ln  V_{H2}(\xi_2; \vec{{\rm x}}_2)\,.
\label{1.7} 
\ee
Here  we use that the ``heavy'' operators carry large charges, so that
  $\ln V_{H1,H_2}$,  like $I$, are proportional to string tension. 
%
Then  
\be
G_3(\vec{{\rm x}}_1, \vec{{\rm x}}_2, \vec{{\rm x}}_3) =
\int d^2 \xi\ V_L (\xi; \vec{{\rm x}}_3)
\int d^2 \xi_1 d^2 \xi_2\ e^{-I} \  V_{H1}(\xi_1; \vec{{\rm x}}_1) V_{H2}(\xi_2; \vec{{\rm x}}_2)\,,
\label{1.8}
\ee
where $I, V_H, V_L$ are now  evaluated on the solution to the equations 
of motion that follow from \eqref{1.7}.
The second factor in~\eqref{1.8} is the 
semiclassical value of the two-point function of the two ``heavy'' operators. 
If we divide by it we end up with 
\be 
\frac{G_3 (\vec{{\rm x}}_1, \vec{{\rm x}}_2, \vec{{\rm x}}_3)}{G_2(\vec{{\rm x}}_1, \vec{{\rm x}}_2)}
=\int d^2 \xi\ V_L (z(\xi), \vec{x}(\xi)-\vec{{\rm x}}_3, X_k (\xi))\,,
\label{1.9}
\ee
where $(z(\xi), \vec{x}(\xi), X_k(\xi))$  represents 
 the corresponding  string solution 
carrying the same quantum numbers as the ``heavy'' vertex operators
with the boundary conditions~\eqref{1.5} transformed to the $\xi$-plane 
using~\eqref{1.2}.  Using 2d  conformal invariance we can also  transform~\eqref{1.9} 
back to the cylinder to get ($\int d^2 \sigma = \int^\infty_{-\infty}
 d \tau_e \int^{2\pi}_0 d \sigma$)
\be
\frac{G_3 (\vec{{\rm x}}_1, \vec{{\rm x}}_2,
 \vec{{\rm x}}_3)}{G_2(\vec{{\rm x}}_1, \vec{{\rm x}}_2)}=
\int d^2 \sigma \ 
V_L (z(\tau_e, \sigma), \vec{x}(\tau_e, \sigma)-\vec{{\rm x}}_3, X_k(\tau_e, \s))\,.
\label{1.10}
\ee
%
This expression captures the leading dependence on $\sqrt{\lambda}\gg 1$
(the validity of this approximation was  discussed in detail 
in~\cite{rt}).
The global  conformal $SO(2,4)$ symmetry fixes the form of 
  the  two-point and three-point functions
 (we assume  that the  operators  correspond to scalar 
 primaries and $\V_{H2}= \V^*_{H1}$) 
%
\bea 
&&
  G_2  (\vec{{\rm x}}_1, \vec{{\rm x}}_2)=
\frac{C_{12}\ \delta_{\Delta_1, \Delta_2} }{{\rm x}_{12}^{\Delta_1+\Delta_2}}
\ , \ \ \ \ \ \ \ \ \ \ \ \ \ {\rm x}_{i j}\equiv |\vec{{\rm x}}_i -\vec{{\rm x}}_j| \ , 
\la{1.01}\\
&& G_3  (\vec{{\rm x}}_1, \vec{{\rm x}}_2, \vec{{\rm x}}_3)=
\frac{C_{123}}{{\rm x}_{12}^{\Delta_1+\Delta_2-\Delta_3}
{\rm x}_{13}^{\Delta_1+\Delta_3-\Delta_2}
{\rm x}_{23}^{\Delta_2+\Delta_3-\Delta_1}}\,,
\label{1.11}
\eea
where 
$\Delta_i$ are the dimensions 
of the operators. By choosing the locations $\rm x_i$ appropriately one can remove the 
dependence on ${\rm x}_{i j}$ in~\eqref{1.10} and 
adapt~\eqref{1.10} to computing the 
coefficient $C_{123}$~\cite{z, rt}.  
Assuming that  $\Delta_1=\Delta_2$  (as their possible  difference is subleading 
in the approximation we consider) 
 we then find  (choosing ${\rm x}_3=0$)\foot{Here  we formally set $C_{12}=1$ in \rf{1.01}, i.e.
 assumed  that the ``heavy'' operators  are normalized.
 The ratio  $G_3/G_2$  does not, of course,
 depend on the normalization of the ``heavy'' operators, i.e. 
 what we  will be computing  below is, in fact, the invariant ratio $C_{123}/C_{12}$.}
\be 
\frac{G_3 (\vec{{\rm x}}_1, \vec{{\rm x}}_2,
 \vec{{\rm x}}_3=0)}{G_2(\vec{{\rm x}}_1, \vec{{\rm x}}_2)}=
C_{123} \ \Big( \frac{ \rx_{12}}{ |\rx_1| \  | \rx_2|}\Big)^{\Delta_3}
 \ . \la{js}
 \ee
As was suggested in \ci{rt}, the  same logic   can be applied to 
semiclassical   computation of  any
 $n$-point correlation function  that contains 
  two ``heavy'' and $n$  ``light''   operators.
  Here we shall  focus  on  the  case of the four-point correlator 
%
\bea
&&G_4 (\vec{{\rm x}}_1, \vec{{\rm x}}_2, \vec{{\rm x}}_3, \vec{{\rm x}}_4)=
\langle \V_{H1}(\vec{{\rm x}}_1) \V_{H2}(\vec{{\rm x}}_2)\V_{L1}(\vec{{\rm x}}_3)\V_{L2}(\vec{{\rm x}}_4) \rangle
\label{1.12}\\
&&=\int {\cal D} {\mathbb X}^{\mathbb M}\ e^{-I}\int d^2 \xi_1 d^2 \xi_2d^2 \xi_3 d^2 \xi_4
\ V_{H1}(\xi_1; \vec{{\rm x}}_1) V_{H2}(\xi_2; \vec{{\rm x}}_2) 
V_{L1}(\xi_3; \vec{{\rm x}}_3) V_{L2}(\xi_4; \vec{{\rm x}}_4)\,.
\nonumber
\eea
The semiclassical trajectory is  again the same, i.e. is 
obtained from  ~\eqref{1.7}, 
and to compute the leading semiclassical  term in $G_4$ we need to 
evaluate the action $I$  and the product of ``light'' operators on this solution
\bea  &&
\frac{G_4(\vec{{\rm x}}_1, \vec{{\rm x}}_2, \vec{{\rm x}}_3, 
\vec{{\rm x}}_4)}{G_2(\vec{{\rm x}}_1, \vec{{\rm x}}_2)} \label{1.13}
\\
&&
 =
\int d^2 \xi_3\ V_{L1}(z(\xi_3), \vec{x}(\xi_3)-\vec{{\rm x}}_3, X_k(\xi)) 
\int d^2 \xi_4\ V_{L2}(z(\xi_4), \vec{x}(\xi_4)-\vec{{\rm x}}_4, X_k(\xi))
\nonumber 
\eea
where we divided by the two-point function of the ``heavy'' operators as in~\eqref{1.9}.
Note that the integrals over $\xi_3$ and $\xi_4$ decouple from each other, i.e. 
the four-point function factorizes. 
Transforming to the  $(\tau_e, \sigma)$-coordinates we get
\bea
 \frac{G_4(\vec{{\rm x}}_1, \vec{{\rm x}}_2, \vec{{\rm x}}_3, \vec{{\rm x}}_4)}
 {G_2(\vec{{\rm x}}_1, \vec{{\rm x}}_2)}  = & &
\int d^2 \sigma\ V_{L1}(z(\tau_e, \sigma), \vec{x}(\tau_e,
 \sigma)-\vec{{\rm x}}_3, X_k(\tau_e, \s)) 
\nonumber\\
&\times&  \int d^2 \sigma'\ V_{L2}(z(\tau'_e, \sigma'), \vec{x}(\tau'_e, \sigma')-\vec{{\rm x}}_4,
 X_k(\tau'_e, \s'))\,.
\label{1.14} 
\eea
%
According to~\eqref{1.10} each integral in~\eqref{1.14} is the ratio of the three- and
two-point functions. Then we obtain the following factorization
\bea
 \langle \V_{H1}(\vec{{\rm x}}_1) \V_{H2}(\vec{{\rm x}}_2)
\V_{L1}(\vec{{\rm x}}_3)\V_{L2}(\vec{{\rm x}}_4) \rangle &= &
\frac{  \langle \V_{H1}(\vec{{\rm x}}_1) \V_{H2}(\vec{{\rm x}}_2)
\V_{L1}(\vec{{\rm x}}_3) \rangle\ 
 \langle \V_{H1}(\vec{{\rm x}}_1) \V_{H2}(\vec{{\rm x}}_2)
\V_{L2}(\vec{{\rm x}}_4) \rangle }
{ \langle \V_{H1}(\vec{{\rm x}}_1) \V_{H2}(\vec{{\rm x}}_2) \rangle} \no \\ 
 G_4(\vec{{\rm x}}_1, \vec{{\rm x}}_2, \vec{{\rm x}}_3, \vec{{\rm x}}_4) &=&
\frac{G_3(\vec{{\rm x}}_1, \vec{{\rm x}}_2, \vec{{\rm x}}_3)\ G_3(\vec{{\rm x}}_1, \vec{{\rm x}}_2, 
\vec{{\rm x}}_4)}{G_2(\vec{{\rm x}}_1, \vec{{\rm x}}_2)}\,
\label{1.15}\ . \eea
This relation has an obvious generalisation to the case of correlators with two ``heavy'' and
 many  ``light'' vertex operators.

Let us finish this section with a few comments. 
The semiclassical expression  \rf{1.14},\rf{1.15} 
for the above 4-point function  may be interpreted as describing the
  process in which a ``heavy''
classical  
   string emits  two  ``light''  quantum strings, each at separate time.
   The above  semiclassical path integral  argument is already 
   a sufficient justification that this  is a dominant process at 
   large $\lambda$. In fact,  the  process  in which  the
 ``macroscopic'' string first emits 
one ``light'' string mode   which then  decays into two other ``light'' modes
is  subleading at large $\sql$. 
As we shall see  explicitly in the examples considered below, each of the (normalized)
3-point functions in  \rf{1.15} will    scale as $\sql \gg 1 $  while a correlator  
of  3 ``light'' states is of  order 1.

As is well known, 
unlike two- and three-point correlators, 
the $\rm x_i$   dependence of the 
four-point correlators  is  not fixed by the  conformal invariance:
in general, they involve  non-trivial 
functions of the conformal cross ratios
\be
u=\frac{{\rm x}_{12}^2 {\rm x}_{34}^2}{{\rm x}_{13}^2 {\rm x}_{24}^2}\,, \qquad \qquad
v=\frac{{\rm x}_{14}^2 {\rm x}_{23}^2}{{\rm x}_{13}^2 {\rm x}_{24}^2}\,.
\label{1.16}
\ee
The factorization \rf{1.15}  predicted  for the leading term in the semiclassical expansion 
of $\langle V_{H1}(\vec{{\rm x}}_1) V_{H2}(\vec{{\rm x}}_2)
V_{L1}(\vec{{\rm x}}_3)V_{L2}(\vec{{\rm x}}_4) \rangle$ 
implies a particular dependence on conformal cross-ratios. 
Each of the three-point functions in~\eqref{1.15} 
is of the form $\langle V_H V_H V_L\rangle$  and its form is fixed by 
\rf{1.11}  but the precise quantum numbers 
of the ``heavy'' operators in~\eqref{1.15}, 
 in general, 
may  not coincide with the quantum numbers of the original operators in
$G_4(\vec{{\rm x}}_1, \vec{{\rm x}}_2, \vec{{\rm x}}_3, \vec{{\rm x}}_4)$
as they  may  be shifted by the dimensions of the ``light'' operators. 
This shift is important because $\langle V_H V_H V_L\rangle$ depends, in particular, 
on the difference of the dimensions of ``heavy'' operators (see eq.~\eqref{1.11})
in which such a shift can produce  a leading order contribution.
We will see an example of this in Section 4.


\section{4-point correlator with  two dilaton operators}


Let us now follow the discussion in \ci{rt} and 
 provide a consistency check of the factorization ~\eqref{1.15}
by taking the two ``light'' operators to be the dilaton operators (with zero $S^5$ momentum). 
In this case it is possible to 
give an independent derivation of the expression in ~\eqref{1.15}.

The vertex operator for the dilaton inserted at 
the point $\vec{{\rm x}}$ on the boundary
is given by (see, e.g., \ci{rt} and refs. there)
\be
\V_{dil}(\vec{{\rm x}})=
c_{dil}\int d^2 \xi\   K_4(\xi;\vec{{\rm x}}) \   L  \,,
 \ \ \ \ \ \ \ \ K_4(\xi;\vec{{\rm x}})=  
 \Big(\frac{z}{z^2 + (\vec{x}-\vec{{\rm x}})^2}\Big)^4 \ , 
\label{2.1}
\ee
where $c_{dil}$ is a normalization coefficient and
$L$ is the string Lagrangian in \rf{st}. 
If we integrate $\V_{dil}(\vec{{\rm x}})$ over the Euclidean 4-space 
 $\vec{{\rm x}}$
the factor $K_4$   goes away  and we end up with  the 
expression proportional to string action 
%
%
%
\be
\int d^4 {\rm x}\  \V_{dil}(\vec{{\rm x}}) =  a_{dil}  \int d^2\xi\  L\,, \ \ \ \
\ \ \  \   a_{dil}= { \pi^2 \ov 6}  c_{dil} \ . 
\label{2.2}
\ee
The 
proportionality coefficient here  
  is  independent of $\vec{x}$ and 
$z$  (this  is easy to see by 
first translating $\vec{{\rm x}}$ by $\vec{x}$, then rescaling
 $\vec{{\rm x}}$ by $z$ and finally doing the integral). 

Let  us  now  consider the general expression for the 
three-point function involving two ``heavy'' operators of dimension 
$\Delta$ ($\V_{H1}= \V_{H2}^*$)  and the dilaton, 
\be 
\langle \V_{H1}(\vec{{\rm x}}_1)\V_{H2}(\vec{{\rm x}}_2) \V_{dil}(\vec{{\rm x}}_3)\rangle=
\frac{C_{\Delta, dil}}{{\rm x}_{12}^{2 \Delta -4} {\rm x}_{13}^4 {\rm x}_{23}^4}\,,
\label{2.3}
\ee
and integrate it over $\vec{{\rm x}}_3$. The l.h.s. of \rf{2.3}  then gives
\be
\langle \V_{H1}(\vec{{\rm x}}_1)\V_{H2}(\vec{{\rm x}}_2) \int
 d^4 {\rm x}_3\
\V_{dil}(\vec{{\rm x}}_3)\rangle  = a_{dil}  
\langle \V_{H1}(\vec{{\rm x}}_1)\V_{H2}(\vec{{\rm x}}_2)\  \int d^2 \xi\  L\rangle \,.
\label{2.3.1}
\ee
Since the  averaging is done with the  measure $e^{-I}$ 
where $I=   \int d^2 \xi\  L$ is the string action \rf{st} 
containing a factor of $\sql$ and  thus 
satisfying $ \lambda \frac{\partial}{\partial \lambda} I = \ha   I$
the r.h.s. of   \rf{2.3.1}   may   be written  also as 
%
\be
\lambda \frac{\partial}{\partial \lambda}\langle \V_{H1}(\vec{{\rm x}}_1)
\V_{H2}(\vec{{\rm x}}_2) \rangle 
\ \ \sim \ \  \lambda \frac{\partial \Delta}{\partial \lambda}
\frac{\ln ( {\rm x}_{12}^2 \mu^2) }{{\rm x}_{12}^{2 \Delta}}\,, 
\label{2.4}
\ee
where in the  r.h.s.   we  used \rf{1.11}  ($\mu$ is a  normalization scale or a cutoff). 
At the same time, integrating the r.h.s.  of~\eqref{2.3} gives
\be 
\frac{C_{\Delta, dil}}{{\rm x}_{12}^{2 \Delta-4}}\int \frac{d^4 {\rm x}_3}{{\rm x}_{13}^4
 {\rm x}_{23}^4} 
\ \sim \  C_{\Delta, dil} \frac{\ln ( {\rm x}_{12}^2\mu^2) }{{\rm x}_{12}^{2 \Delta}}\,.
\label{2.5}
\ee
Comparing~\eqref{2.4} and~\eqref{2.5} we finish with  the following relation  
(see also \ci{cost,rt})\foot{Similar argument leading to this relation 
can be given on the gauge theory side where $\l^{-1} 
$ appears as a coefficient in from the
action and the integrated dilaton operator is 
proportional to the gauge theory  action.} 
\be 
C_{\Delta, dil} \ \ \sim\ \  \lambda \frac{\partial \Delta}{\partial \lambda}\,.
\label{2.6}
\ee
Let  us now  insert one more dilaton operator  and integrate over its position:
\bea
&&\int d^4 {\rm x}_4\ \langle \V_{H1}(\vec{{\rm x}}_1)\V_{H2}(\vec{{\rm x}}_2) 
\V_{dil} (\vec{{\rm x}}_3) \V_{dil}(\vec{{\rm x}}_4)\rangle
\sim \langle \V_{H1}(\vec{{\rm x}}_1)\V_{H2}(\vec{{\rm x}}_2) \V_{dil} (\vec{{\rm x}}_3) \int d^2
\xi\ L\rangle
\nonumber \\
&& \qquad \qquad \qquad  \qquad  \sim\ \lambda \frac{\partial}{\partial \lambda}
\langle V_{H1}(\vec{{\rm x}}_1)V_{H2}(\vec{{\rm x}}_2) V_{dil} (\vec{{\rm x}}_3) \rangle\,.
\label{2.7}
\eea
Using that the 
 three-point function $\langle \V_{H1}(\vec{{\rm x}}_1)\V_{H2}(\vec{{\rm x}}_2)
  \V_{dil} (\vec{{\rm x}}_3) \rangle$
is determined  by ~\eqref{2.3}, \eqref{2.6}   we get 
\be
\int d^4 {\rm x}_4\ \langle \V_{H1}(\vec{{\rm x}}_1)\V_{H2}(\vec{{\rm x}}_2) 
\V_{dil} (\vec{{\rm x}}_3) \V_{dil}(\vec{{\rm x}}_4)\rangle\sim
\lambda \frac{\partial}{\partial \lambda} 
\bigg( \frac{ \lambda \frac{\partial \Delta}{\partial \lambda}}{{\rm x}_{12}^{2 \Delta -4} {\rm x}_{13}^4 {\rm x}_{23}^4}
\bigg)\,.
\label{2.8}
\ee
Differentiating  the bracket we  find two terms. The first one comes 
from differentiating the numerator and is proportional to 
$ \lambda \frac{\partial}{\partial \lambda}  (\lambda \frac{\partial \Delta}{\partial \lambda})$.
The second term comes from differentiating $\Delta$ in the denominator and is of order
$ (\lambda \frac{\partial \Delta}{\partial \lambda})^2$. In the semiclassical limit 
of large $\Delta$  (scaling as  $\l^{1/2}$)
the first term can be ignored and so  we get \ci{rt}
\be
\int d^4 {\rm x}_4\ \langle \V_{H1}(\vec{{\rm x}}_1)\V_{H2}(\vec{{\rm x}}_2) 
\V_{dil} (\vec{{\rm x}}_3) \V_{dil}(\vec{{\rm x}}_4)\rangle\ \ \sim\ \ 
\Big( \lambda \frac{\partial \Delta}{\partial \lambda}\Big)^2\ 
\frac{\ln ({\rm x}_{12}^2\mu^2)}{{\rm x}_{12}^{2 \Delta}} \frac{1}{{\rm x}_{13}^4 {\rm x}_{23}^4}\,.
\label{2.9}
\ee
From our discussion of the three-point function earlier
 in this section we already know 
that this implies that
\be
\langle \V_{H1}(\vec{{\rm x}}_1)\V_{H2}(\vec{{\rm x}}_2) \V_{dil} (\vec{{\rm x}}_3) 
\V_{dil}(\vec{{\rm x}}_4)\rangle \ \ \sim\ \ 
\frac{\lambda \frac{\partial \Delta}{\partial \lambda}}{{\rm x}_{12}^{2 \Delta}{\rm x}_{14}^4 
{\rm x}_{24}^4}
\   
\frac{\lambda \frac{\partial \Delta}{\partial \lambda}}{{\rm x}_{12}^{2 \Delta}{\rm x}_{13}^4
 {\rm x}_{23}^4}
\ \ {\rm x}_{12}^{2 \Delta}\,.
\label{2.10}
\ee
This is precisely the factorized expression in \eqref{1.15} 
\be 
\langle \V_{H1}(\vec{{\rm x}}_1)\V_{H2}(\vec{{\rm x}}_2) \V_{dil} (\vec{{\rm x}}_3)
 \V_{dil}(\vec{{\rm x}}_4)\rangle\ =
\frac{\langle \V_{H1}(\vec{{\rm x}}_1) \V_{H2}(\vec{{\rm x}}_2) \V_{dil}(\vec{{\rm x}}_3)\rangle\ 
\langle \V_{H1}(\vec{{\rm x}}_1) \V_{H2}(\vec{{\rm x}}_2) \V_{dil}(\vec{{\rm x}}_4)\rangle}{\langle
\V_{H1}(\vec{{\rm x}}_1) \V_{H2}(\vec{{\rm x}}_2\rangle}\,.
\label{2.11}
\ee
We have thus independently proven  ~\eqref{1.15}
 in the case when the two ``light'' operators  are the 
dilaton ones. This provides  a non-trivial consistency check of the  
general semiclassical prediction~\eqref{1.15} as was already 
mentioned in \ci{rt}.\foot{As was  discussed  at the end of section 2, 
the contribution to the  above 4-point correlator corresponding to  the ``heavy'' state
emitting a ``light'' state (a ``graviton'') 
 that then decays into 2 dilatons should be subleading at
large $\l$.  Indeed,  such contribution may come from the region of the integral  
over ${\rm x}_4$ where  ${\rm x}_4$ approaches  ${\rm x}_3$  and would then 
be proportional  to the  product of 3-point function with two  ``heavy'' states and 
one ``light''  state  and three ``light'' states. As the former will scale as $\sql$
while the  latter  will be of order 1, this contribution will   be negligible compared to
\rf{2.9}  which scales as $(\sql)^2$    for $\Delta \sim \sql$.}

The  above argument can be easily   generalized to the case 
of more than two dilaton operator insertions. In this case the dominant contribution to the
relation like  \rf{2.8}   will be coming  again  from the term with  maximal power of 
$\lambda \frac{\partial \Delta}{\partial \lambda}$  which will dominate at strong coupling
over terms with multiple derivatives of $\Delta$. 


\section{Explicit form of  semiclassical 
correlators involving chiral primary  and 
   twist-two operators }

In this section we will  find    explicit form of some four-point correlation 
functions in the leading semiclassical limit.
We will start with the case of four chiral primary operators (CPO's). To understand the 
factorization \eqref{1.15} in detail we need to revisit their three-point 
function~\cite{z, rt}.


\subsection{3-point function of chiral primary operators revisited}

Let us consider the three-point function
\be
G_3(\vec{{\rm x}}_1, \vec{{\rm x}}_2, \vec{{\rm x}}_3) =\langle 
\V_{-J}(\vec{{\rm x}}_1)\V_{J-j}(\vec{{\rm x}}_2)\V_j(\vec{{\rm x}}_3)
\rangle \,, \qquad\ \ \  J \gg j\,.
\label{3.1}
\ee
Below we shall   label vertex operators by their charges (or spins) rather than dimensions. 
Here $\V_j$ stands  for a ``light'' 
 chiral primary operator   with fixed charge ($S^5$   angular momentum) $j$
while   $\V_J$  is its ``heavy'' counterpart with   large  charge  $J \sim \sql.$
In the leading  semiclassical approximation we may assume 
that $\V_{J-j}\approx V_J$
but it will be useful to keep this distinction 
(and thus have manifest charge conservation)
in a part of the discussion that follows.

According to~\eqref{1.10} in the limit 
of large $J$ this three-point function is given by
\be
\frac{G_3(\vec{{\rm x}}_1, \vec{{\rm x}}_2, \vec{{\rm x}}_3)}{G_2(\vec{{\rm x}}_1,
 \vec{{\rm x}}_2)}=
\int d^2 \sigma\  V_{j} \big(z(\tau_e, \s), \vec{x}(\tau_e, \s)-\vec{{\rm x}}_3, X_k(\tau_e,
 \s)\big)\,.
\label{3.2}
\ee
%
Here  $z(\tau_e, \s)$, $\vec{x}(\tau_e, \s)$ and $X_k(\tau_e, \s)$ correspond to
a point-like string  orbiting big circle of  $ S^5$; the corresponding  Euclidean trajectory in 
the Poincar\'e patch of $AdS_5$ 
satisfies the  boundary conditions~\eqref{1.5}.
Since the three-point function should depend
 only on the absolute values of the coordinate
differences,
without loss of generality we can choose all the points to lie along the ${\rm x}_{0e}$-axis.
Let us denote the zeroth components of $\vec{{\rm x}}_1$ and $\vec{{\rm x}}_2$ as
$a_1$ and $a_{2}$ respectively. For concreteness we will assume that 
$a_{1}>a_{2}$.
Then the corresponding stationary-point 
solution is given by~\cite{bt1} ($a_{12}\equiv  a_{1} - a_{2}$)
\bea
&& z=\frac{a_{12}}{2 \cosh(\kappa \tau_e)}\,, \qquad \ \ \ \ 
x_{0e}=\frac{a_{12}}{2} \tanh (\kappa \tau_e) + \frac{1}{2} (a_1+a_2)\,,
\nonumber\\
&& x_1=x_2=x_3=0\,, \nonumber\\
&& \phi=-i \nu \tau_e\,,  \ \ \  \ \ \ \ J= \sql  \nu  \ , \ \ \ \ \  \  \k = \nu \
, 
\label{3.3}
\eea
where $\phi$ is the angle of $S^5$. 

The expression for the 
``light'' chiral primary vertex operator can be obtained 
starting from the  general  expression 
for the $SO(2,4)$ covariant 10-d graviton ``wave function'' \ci{z,ber,Lee}.
As we show in Appendix A, it can be put in the following form 
(for simplicity, we choose  the  location of the operator to be at 
$\vec{{\rm x}}_3=0$,  but the dependence on $\vec{{\rm x}}_3$ 
can  be  easy to restore at the end)\foot{Here we normalize 
the chiral primary scalar operator  as in \ci{z,ber}.
$\sql \ov 2\pi $ factor in $\cc_j$ is the   string tension (the graviton operator is a
perturbation of the graviton coupling term 
in the string action)  while  $1/N$    stands for
a formal  factor of string coupling ($N\gg 1$ 
is the rank of the gauge group of the dual gauge theory with $1/N$  being the  
standard  normalization of planar 3-point functions).}
\bea
&&\V_j(0)  = \cc_j \int d^2 \s \  K_j\  \rX^j \ U \ ,
 \label{3.4}\\
&& 
K_j = \Big(\frac{z}{z^2 + \vec{x}^2}\Big)^j , \ \ \ \ \ \   \ \ \ \ 
 \rX \equiv X_1 + i X_2 \ , \ \ \ \ \ \ \ \ \ \ 
\cc_j= { 1  \ov N}{\sql \ov 8 \pi}  { (j+1) \sqrt{j} }
 \ . \la{ccc}
\eea
The operator $U$ here  has the following structure
\bea
&&  U=U_1+U_2+U_3 \ , \label{3.14}\\
&& U_1= \frac{1}{z^2}
\Big[(\partial_{\alpha} \vec{x})^2- (\partial_{\alpha} z)^2\Big]-
(\partial_{\alpha} X_k)^2\,, \label{3.6} \\
&&
U_2 =  \frac{8}{(z^2+\vec{x}^2)^2}
\Big[\vec{x}^2 (\partial_{\alpha}z)^2  - (\vec{x} \cdot \partial_{\alpha}\vec{x})^2 \Big]\,, 
 \ \ \  
 U_3 = \frac{8(\vec{x}^2-z^2)}{z( z^2 +\vec{x}^2)^2}(\vec{x} \cdot \partial_{\alpha}\vec{x})
\partial_{\alpha} z\,.
\label{3.15}
\eea
Let us  now  evaluate \eqref{3.4} 
on the solution~\eqref{3.3}. We  get    $\rX^j
  = e^{i j \phi}= e^{j \kappa \tau_e}$  and  (see Appendix A for details)
\bea
U_1 =\frac{2  \kappa^2}{\cosh^2 
\kappa \tau_e}\,,
\ \ \ \ \ \ \  U_2 +U_3 =-\frac{2  \kappa^2}{\cosh^2 \kappa \tau_e}
\frac{(a_{1}^2-a_{2}^2) (a_{1}^2 
e^{2 \kappa \tau_e} -a_{2}^2 e^{-2 \kappa \tau_e})}
{(a_{1}^2 e^{\kappa \tau_e}+ a_{2}^2 e^{-\kappa \tau_e})^2}\,.
\label{3.16}
\eea
Substituting  this into~\eqref{3.4} 
we obtain
\be
\frac{G_3(\vec{{\rm x}}_1, \vec{{\rm x}}_2, \vec{{\rm x}}_3=0)}{G_2(\vec{{\rm x}}_1, 
\vec{{\rm x}}_2)}=
 16 \pi  \kappa^2\  \cc_j  \int_{-\infty}^{\infty} d \tau_e\
\frac{a_{1}^2\ a_{2}^2\ a_{12}^j\ e^{j \kappa \tau_e}}
{(a_{1}^2 e^{\kappa \tau_e} +a_{2}^2 e^{-\kappa \tau_e})^{j+2}}\,.
\label{3.17}
\ee
After performing the integral (first 
rescaling $\k \tau_e\to \tau_e$  and then 
shifting $ \tau_e\to \tau_e + \ln { a_{2} \ov a_{1}}$)  we find 
\be
G_3(\vec{{\rm x}}_1, \vec{{\rm x}}_2, \vec{{\rm x}}_3=0)= 
\frac{C_{J, j}}{{\rm x}_{12}^{2 J-j}\ |\vec{{\rm x}}_{1}|^{2 j}}\approx
\frac{C_{J, j}}{{\rm x}_{12}^{2 J}\ |\vec{{\rm x}}_{1}|^{2 j}}\,,
\label{3.18}
\ee
where
\be
C_{J, j}=\cc_j    \frac{ 8 \pi  \kappa}{j+1}= \frac{1}{N} J \sqrt{j}\ .
\label{3.19}
\ee
We have restored the Lorentz invariance by
replacing $a_{1}$ and $a_{12}$ with $|\vec{{\rm x}}_{1}|$ and 
$\rx_{12}$ (recall that in~\eqref{1.11} we defined
${\rm x}_{i j}\equiv |\vec{{\rm x}}_i -\vec{{\rm x}}_j|$).
To restore the dependence on non-zero 
$\vec{{\rm x}}_3$ in the right hand side of \eqref{3.18} 
we may  simply replace $|\vec{{\rm x}}_1|$ with ${\rm x}_{13}$.

Note that the dependence on ${\rm x}_{ij}$ in \rf{3.18} 
came out to be consistent with \rf{1.11}  with  
\be 
\Delta_1+\Delta_2 \simeq 2J \,, \  \ \ \ \ \ \ \  \Delta_1-\Delta_2=\Delta_3 =j 
\label{3.19.0}
\ee
which implies, for $J \gg j$, that 
\be
\Delta_1=J\,, \ \ \quad \Delta_2 =J-j\,, \ \ \quad \Delta_3=j
\label{3.19.1}
\ee
and is thus  consistent with the charge conservation. 

Note that  if we 
took the charge of the ``light'' operator to be $-j$ instead of $j$ 
we would  get a similar expression  with the same  $C_{J, j}$
and 
${\rm x}_1$ and ${\rm x}_2$ interchanged
\be
G_3(\vec{{\rm x}}_1, \vec{{\rm x}}_2, \vec{{\rm x}}_3=0)= 
\frac{C_{J, j}}{{\rm x}_{12}^{2 J}\ |\vec{{\rm x}}_{2}|^{2 j}} \ , 
\label{3.20}
\ee
with the ${\rm x}_3$-dependence
restored again  by 
$|\vec{{\rm x}}_2| \to {\rm x}_{23}$.
The latter expression 
corresponds to the three-point function of operators of dimensions
$
\Delta_1=J\,, \ \Delta_2 =J+j\,, \ \Delta_3=j\ , $
%
i.e. with the  charges $-J, J+j, -j$.
These  observations about the precise charges of the 
participating chiral primary operators 
 will be 
important in the next subsection where  we will discuss their four-point function. 


It is useful to note that 
 $ U_2 +U_3 $ in \rf{3.16} 
vanishes if $a_1^2-a_2^2=0$, i.e. for the choice of 
$a_1=-a_2$. Computing the 3-point correlator \rf{3.2} 
in the particular case of ${\rm x}_1= - {\rm x}_2$ 
one  can then  use the ``simplified''
version of the vertex operator \rf{3.4}  with $U$ given just by 
$U_1$ in \rf{3.6}. This  operator was used 
in~\cite{z, rt}.\foot{In \ci{z} the justification for its use
instead of the full vertex of \ci{ber,Lee}  was  based 
on  considering the special  case when the  insertion point 
$\vec{\rx}_3$ is sent to $\infty$ 
(which  is, in general,  enough due to conformal invariance). 
In this case the  addition terms  $U_2,U_3$ in \rf{3.15}
are suppressed. 
In \ci{rt}  the  three-point correlator was  chosen to have 
$a_1=-a_2$ and in this case again 
the additional contributions vanish as was explained above.} 

\subsection{4-point function of two ``heavy'' and two ``light''\\ chiral primary operators 
 }

Let us now  apply the results of the previous section to
compute  the four-point function 
of the  chiral primary operators  carrying charges 
$(-J, J, -j, j)$ in the limit of large $J\gg  j $.
According to the semiclassical  prescription~\eqref{1.13} it is given by 
\be
\frac{G_4(\vec{{\rm x}}_1, \vec{{\rm x}}_2, \vec{{\rm x}}_3,
 \vec{{\rm x}}_4)}{G_2 (\vec{{\rm x}}_1,\vec{{\rm x}}_2)}=
\V_{-j}(\vec{{\rm x}}_3) \cdot \V_{j}(\vec{{\rm x}}_4)\,,
\label{3.20.2}
\ee
where $V_{\mp j}(\vec{{\rm x}}_{3,4})$ are 
 given by \eqref{3.4} with $\vec{x}$ shifted 
by $\vec{{\rm x}}_{3, 4}$ and  understood to be 
evaluated on the solution~\eqref{3.3}.
All the necessary ingredients were found  in the previous subsection 
and now we only need to collect them to arrive at the result
\be
G_4(\vec{{\rm x}}_1, \vec{{\rm x}}_2, \vec{{\rm x}}_3, \vec{{\rm x}}_4)=
\frac{
\langle \V_{-J}(\vec{{\rm x}}_1)\V_{J+j}(\vec{{\rm x}}_2)\V_{-j}(\vec{{\rm x}}_3)\rangle
\ \langle \V_{-J}(\vec{{\rm x}}_1)\V_{J-j}(\vec{{\rm x}}_2)\V_{j}(\vec{{\rm x}}_4)\rangle}
{\langle \V_{-J}(\vec{{\rm x}}_1)\V_{J}(\vec{{\rm x}}_2)\rangle}\,.
\label{3.21}
\ee
Note that the charges of the ``heavy'' operators appearing in the 
three-point functions in~\eqref{3.21} are formally different from 
their original charges $(-J, J)$, 
in the four-point function. 
As was already mentioned at the end of Section 2, it is important 
to keep track of the precise charges when
writing factorised  expressions in terms of 
 the three-point functions.
The right hand side of eq.~\eqref{3.2} computes the leading contribution 
to the ratio of the three- and two-point functions in the limit of large 
$\lambda$ and $J$. This leading contribution depends not only on the dimensions
of the ``heavy'' operators but also on the difference of their dimensions which is 
sensitive to the shift of $J$ by $\pm j$.
On the other hand, the structure constant $C_{J, j}$ depends only on the dimensions
of the operators and not on their difference. Hence, in computing $C_{J, j}$ 
such a shift is not important. 
Note also that  both three-point functions in \rf{3.21} are consistent with charge
conservation.

Using \rf{3.18}--\rf{3.20}  we may write~\eqref{3.21} in a  more explicit form as
\be
G_4(\vec{{\rm x}}_1, \vec{{\rm x}}_2, \vec{{\rm x}}_3, \vec{{\rm x}}_4)=\frac{j J^2}{N^2} 
\ \frac{1}{{\rm x}_{12}^{2J} \ {\rm x}_{23}^{2j} \ {\rm x}_{14}^{2j}}\,.
\label{3.22}
\ee
It is   convenient  also  to 
 present $G_4(\vec{{\rm x}}_1, \vec{{\rm x}}_2, \vec{{\rm x}}_3, \vec{{\rm x}}_4)$
 in a  form  involving cross-ratios defined in \rf{1.16} 
%
\be
G_{4} (\vec{{\rm x}}_1, \vec{{\rm x}}_2, \vec{{\rm x}}_3, \vec{{\rm x}}_4)=
\frac{1}{{\rm x}_{12}^{2J} \ {\rm x}_{34}^{2 j}}\ {\cal F} (u, v)\,,
\ \ \ \ \ \ \ \ \ \ 
{\cal F} (u, v)=\frac{j J^2}{N^2} \ \frac{u^j}{v^j}\,,\label{3.23}
\ee
%
%
%
where we used that  $J \gg j$.

Let us now consider a slightly different 4-point correlator
of chiral primary  operators  --  with charges 
$(-J-j_1 -j_2, {\ }J, {\ }j_1, {\ }j_2)$, where   $j_1, j_2 \geq 0$.
Like  similar 3-point correlators considered above, this 
 is an extremal correlator as 
$\Delta_{-J-j_1 -j_2}= J +j_1 + j_2 =\Delta_J + \Delta_{j_1}+ \Delta_{j_2}$. 
For such extremal  correlators of BPS operators 
with  $\Delta_1=\Delta_2+ \dots+ \Delta_n$
 there exists a
non-renormalization conjecture~\cite{106,Freedman}
 (here  for generality we label operators by their dimensions
  and consider only planar
 approximation) 
%
\be
\langle {\V}_{\Delta_1} (\vec{{\rm x}}_1)
{\V}_{\Delta_2} (\vec{{\rm x}}_2)
\dots {\V}_{\Delta_n} (\vec{{\rm x}}_n)
\rangle = { A(\{\Delta_i\})\ov N^{n-2} } \prod_{k=2}^n \frac{1}{{\rm x}_{1k}^{2 \Delta_k}}\,,
\label{3.25}
\ee
where 
the coefficient 
$A(\{\Delta_i\})$ should  not depend on the 't Hooft coupling  $\l$, i.e.  
 should be the same at weak and strong coupling. 
Accoding to the   semiclassical 
prescription~\eqref{1.13} for the above   choice of charges 
of chiral primary operators 
we again obtain the expression in 
eq.~\eqref{3.20.2}. Since both $j_1$ and $j_2$ are  assumed to be positive 
we then find  using~\eqref{3.18} and \eqref{3.19}
\be
G_4(\vec{{\rm x}}_1, \vec{{\rm x}}_2, \vec{{\rm x}}_3, \vec{{\rm x}}_4)=
\frac{J^2 \sqrt{j_1 j_2}}{N^2}\frac{1}{{\rm x}_{12}^{2J}\ {\rm x}_{13}^{2 j_1}
\ {\rm x}_{14}^{2 j_2}}\,.
\label{3.26}
\ee
This expression is in perfect agreement with~\eqref{3.25}
which is not too surprising as this 4-point 
correlator is expressed in terms of extremal 3-point correlators.
 This observation provides another 
consistency check of  the  semiclassical  method
of computing  such   higher-point correlation functions.


\subsection{4-point function of two  large spin $S$  operators \\
and two ``light'' chiral primary operators}

Let us now consider an example of  4-point correlator involving two 
non-BPS ``heavy'' operators 
dual to  large spin   gauge theory operators 
and two ``light'' chiral primary operators.
The minimal twist  large spin   gauge theory operators
are dual  to a folded string with spin  $S$  in $AdS_3$~\cite{gkp2}.
We will denote the corresponding vertex operator as 
$\V_S$, with  $\V_{-S} \equiv  \V^*_S$.
The two-point function of such operators can be computed semiclassically in the 
limit of large $S \gg \sql$) spin~\cite{b1, bt1}
 giving (we assume that $\V_S$ is  normalized appropriately)
\bea
&&\langle \V_{S}(\vec{{\rm x}}_1)\V_{-S} (\vec{{\rm x}}_2)\rangle =\frac{1}{{\rm x}_{12}^{2 \Delta(S)}}\,,
\label{3.3.0} \\
&&\Delta (S)= S+\frac{\sqrt{\lambda}}{\pi} \ln  {S\ov \sql} +\ldots\,. 
\label{3.3.2}
\eea
The corresponding 3-point functions with two large 
spin $S$  and one BPS operator  were already considered in \ci{rt}.
Let us  consider the 3-point function with one ``light'' chiral primary 
operator
\be
G_{3}(\vec{{\rm x}}_1, \vec{{\rm x}}_2, \vec{{\rm x}}_3)=
\langle  \V_{S}(\vec{{\rm x}}_1)\V_{-S} (\vec{{\rm x}}_2) \V_j (\vec{{\rm x}}_3) \rangle\,.
\label{3.3.3}
\ee
For simplicity, we will set
$\vec{{\rm x}}_3=0$. The 3-point function coefficients 
are then determined by   the ratio 
$G_{3}(\vec{{\rm x}}_1, \vec{{\rm x}}_2, \vec{{\rm x}}_3)/G_{2}(\vec{{\rm x}}_1, 
\vec{{\rm x}}_2)$ in 
 eq.~\eqref{3.4}. 
 Compared to \ci{rt} here we 
 start with  generic positions 
 $\vec{\rx}_1,\ \vec{\rx}_2$  and thus need to use the general form 
 of the chiral primary vertex in \rf{3.14}, \eqref{3.15}. 
 The integral in \eqref{3.4} is to be 
 evaluated on the euclidean stationary point  solution 
corresponding to the long folded  spinning string in $AdS_3$ 
\cite{bt1} (cf. \rf{3.3})\footnote{As in the subsection 4.1, we  can choose 
without loss of generality, the points
$\vec{{\rm x}}_1$ and $\vec{{\rm x}}_2$ to lie on the ${\rm x}_{0e}$-axis as 
 the three-point function should depend only  on the absolute 
 values of the 
coordinate differences. 
As above,  we again denote the zeroth components of $\vec{{\rm x}}_1$ and $\vec{{\rm x}}_2$
by $a_1$ and $a_2$.}
\bea
&&z=\frac{a_{12}}{2 \cosh (\kappa \tau_e) \cosh (\mu \s)}\ ,
\ \ \ \ \ \ \ \
x_{0e}= \frac{a_{12}}{2} \tanh (\kappa \tau_e) +\frac{1}{2} (a_1+a_2)\,,
\label{3.3.4}\\
&&x_1 + i x_2 = \frac{a_{12}}{2} \frac{\tanh (\mu \s)}{\cosh (\kappa \tau_e)}\,
e^{ i \varphi} \ , \ \ \ \ \ \ \ \    \varphi=- i   \kappa \tau_e\,,  \ \ \ \ \   x_3 =0 \ , 
\nonumber\\
&&
\kappa =\mu =\frac{1}{\pi} \ln { \S}   \ ,\ \ \ \ \ \ \ \ \ \ \S = { S \ov \sql} \gg 1    \ .  
\label{3.3.4.1}
\eea
This solution~\eqref{3.3.4} is valid in the limit of large $\S$
on the interval $\sigma \in [0, {\pi\ov 2}]$, describing one quarter of the folded string.
If we choose $a_1=-a_2$ then on
this solution $U_2+U_3=0$, i.e.  the  chiral primary vertex 
operator takes its simplified form with $U=U_1$ used in~\cite{rt}. 
Keeping $a_1$ and $a_2$ 
arbitrary we then get from \eqref{3.4}, \rf{3.14}, \eqref{3.15}
(see Appendix A for details)
\bea
&&\frac{G_{3}(\vec{{\rm x}}_1, \vec{{\rm x}}_2, \vec{{\rm x}}_3=0)}{G_{2}(\vec{{\rm x}}_1, \vec{{\rm x}}_2)}=
4 \cc_j  \int_{0}^{\pi/2} d \s \int_{-\infty}^{\infty} d \tau_e \ 
\frac{2 a_{12}^j}{\cosh^j (\mu\s) \ 
(a_1^2 e^{\kappa \tau_e}+a_2^2 e^{-\kappa \tau_e})^j}
\nonumber\\
&&\ \ \ \ \ \ \ \ \ \ \ \ \ \ \ \ \ \ \ \ \ \ \ \ \ \ \ \ \ \ \ 
 \times \Big[ \frac{4\kappa^2 a_1^2 a_2^2 }{(a_1^2 e^{\kappa \tau_e}+a_2^2 e^{-\kappa \tau_e})^2}
-\mu^2 \tanh^2 (\mu \s)\Big]\,.
\label{3.3.5}
\eea
Changing the variables $\kappa \tau_e \to \tau_e, \ \mu \s=\k \s \to \s $ and 
$\tau_e \to \tau_e +\ln {a_2 \ov a_1}$ we can pull all the dependence 
on ${\rm x}_i$ out of the integral to get
\be
G_3(\vec{{\rm x}}_1, \vec{{\rm x}}_2, \vec{{\rm x}}_3=0)=
\frac{C_{S, j}}{{\rm x}_{12}^{2 \Delta(S)}\ |\vec{{\rm x}}_1|^j\ |\vec{{\rm x}}_2|^j}\ ,
\label{3.3.7}
\ee
where 
\be
C_{S, j}=8  \cc_j   \int_{0}^{\ha \pi  \k  } d\s \int_{-\infty}^{\infty} d \tau_e 
\frac{1}{\cosh^j \s \ (e^{\tau_e}+ e^{-\tau_e})^j}
\Big( \frac{1}{\cosh^2 \tau_e}
- \tanh^2  \s \Big)  \,
\label{3.3.8}
\ee
and we have replaced $a_1$, $a_2$ and $a_{12}$ with the Lorentz invariant
objects $|\vec{\rm x}_1|$, $|\vec{\rm x}_2|$ and ${\rm x}_{12}$.
Evaluating the integrals in the limit of large $\kappa$ we 
find that the leading term in $C_{S, j}$  is constant (i.e. 
does not depend on $\k = {1 \ov \pi} \ln \S$)
\be
C_{S, j}\  \approx 
 \  \  \cc_j \ \frac{4 \pi \Gamma[\frac{j}{2}]^2}{2^j \Gamma[\frac{j-1}{2}]
  \Gamma[\frac{j+3}{2}]}
={ 1 \ov N} \
\sqrt{\lambda}\ \frac{\sqrt{j}}{2^j}\  \frac{   \Gamma[\frac{j}{2}]^2}
{\Gamma[\frac{j-1}{2}] \Gamma[\frac{j+1}{2}]}
\,,
\label{3.3.11}
\ee
which is the same as  the expression found in \ci{rt} 
(the leading term in eq.(4.28) there).

To restore the $\vec{{\rm x}}_3$ dependence in \eqref{3.3.7} we  should 
again 
replace ${\rm x}_1$ with ${\rm x}_{13}$ and 
${\rm x}_2$ with ${\rm x}_{23}$.
The structure of  \eqref{3.3.7} is then  the expected one for the  three 
conformal operators   with dimensions 
%
\be
\Delta_1=\Delta_2 = \Delta (S)\,, \ \ \ \ \  \qquad \Delta_3 =j\,.
\label{3.3.9}
\ee
Let us note that so far we ignored the issue of $S^5$ angular momentum conservation 
as its effect is subleading at large $S$. 
We may explicitly 
maintain  the $S^5$  momentum conservation  by considering the operator with 
the charges 
$(S,j_1)$, $(-S, -j-j_1)$, $(0, j)$. The  additional $j, j_1$ dependent 
terms  correcting  \rf{3.3.7},\rf{3.3.11}
will  be suppressed  by a factor
 of $\mu^{-1} \sim (\ln \S)^{-1} \ll 1 $, see \ci{rt,rut}, 
 so that the leading term will not depend on them.


Consider now 
the four-point function\foot{We may assume $j_2=-j_1$ to satisfy explicitly 
the charge
conservation. Then $\rx_{ij}^{j_2} $ below  should  be
 replaced by $\rx_{ij}^{|j_2|} $.}
\be
G_4 (\vec{{\rm x}}_1, \vec{{\rm x}}_2, \vec{{\rm x}}_3, \vec{{\rm x}}_{4})=
\langle
\V_{S}(\vec{{\rm x}}_1)\V_{-S} (\vec{{\rm x}}_2) \V_{j_1} (\vec{{\rm x}}_3) 
\V_{j_2}(\vec{{\rm x}}_4)\rangle\,.
\label{3.3.10}
\ee
According to~\eqref{1.14}  for $S \gg j_1,j_2$ 
it is then given by the product of two chiral primary  vertex
operators evaluated on the solution \eqref{3.3.4}. Using  \rf{3.3.0}  we then obtain
\be
G_4 (\vec{{\rm x}}_1, \vec{{\rm x}}_2, \vec{{\rm x}}_3, \vec{{\rm x}}_{4})=
\frac{C_{S, j_1} \ C_{S, j_2}}{{\rm x}_{12}^{2 \Delta (S)}\ {\rm x}_{13}^{j_1}\ {\rm x}_{23}^{j_1} \
{\rm x}_{14}^{j_2}\ {\rm x}_{24}^{j_2}}\ .
\label{3.3.12}
\ee
We can also present this  in the form of~\eqref{1.15}
\be
G_4 (\vec{{\rm x}}_1, \vec{{\rm x}}_2, \vec{{\rm x}}_3, \vec{{\rm x}}_{4})=
\frac{\langle
\V_{S}(\vec{{\rm x}}_1)\V_{-S} (\vec{{\rm x}}_2) \V_{j_1}(\vec{{\rm x}}_3) \rangle
\langle
\V_{S}(\vec{{\rm x}}_1)\V_{-S} (\vec{{\rm x}}_2) \V_{j_2}(\vec{{\rm x}}_4) \rangle}
{\langle
\V_{S}(\vec{{\rm x}}_1)\V_{-S} (\vec{{\rm x}}_2)\rangle}
\label{ 3.3.13}
\ee
%
Let us   finish with two  comments.

The 3-point coefficient in \rf{3.3.11} is proportional to 
the string tension, i.e.  $\sim \sql \gg 1 $ 
(this factor originates from the  normalization constant in
 the ``light'' vertex operator in 
\rf{ccc} since it is determined by a ``graviton'' perturbation in the string 
action).  The same is also true  for the BPS 3-point  function in \rf{3.19} 
where $J= \sql \k \sim \sql$. As a result, the semiclassical 4-point functions in  
\rf{3.26} or in \rf{3.3.12} scale as  $(\sql)^2$. 

The  expression \rf{3.3.11} implies also  that for  large $j\gg 1 $ 
(but  with   ${ j\ov \sql} \ll 1$ for the  validity of  our approximation)
 one  finds 
 $C_{S, j} \to { \sql \ov N}  \sqrt{j} \ {\rm exp} ( - j \ln 2)$.
 Such exponential suppression is expected for correlators in which all charges of the   
 vertex operators  are large so that  such correlators   should be  dominated 
  by a semiclassical  action  factor $ \sim e^{-  a \sql }$\   (cf. \ci{jan,gro}).
  The same  applies  then also  to the 4-point function \rf{3.3.12}.


\section{Comparison of semiclassical   correlator 
of four chiral primary operators  with  the free gauge theory and \\ supergravity
results}


Let us now compare our  semiclassical result \rf{3.22},\rf{3.23} 
for the  correlator  of two ``heavy'' and two ``light'' chiral primary operators  with 
results available in the literature. 
To be able  to take the semiclassical limit we need a correlator of  BPS 
operators with two arbitrary charges that can  be taken to be large.
As far as we know, the only  such correlator that was considered 
in the literature is that with   two chiral primary operators of
arbitrary  dimension $J$ and two of  dimension 2 \cite{Rayson, Linda1}.

To facilitate   comparison with the expressions in \cite{Rayson, Linda1}
we find it convenient to relabel the coordinates
$
\vec{{\rm x}}_1 \leftrightarrow \vec{{\rm x}}_4\,, \quad  
\vec{{\rm x}}_2 \leftrightarrow \vec{{\rm x}}_3
$
in \eqref{3.23}. Note that the definitions of $u,v$ are kept  the same as in \rf{1.16}.
Then for $j=2$ the expression in \rf{3.23} becomes
\be 
\langle
\V_{2}(\vec{{\rm x}}_1)\ \V_{-2} (\vec{{\rm x}}_2)\ \V_{J} (\vec{{\rm x}}_3)\ 
 \V_{-J}(\vec{{\rm x}}_4)\rangle_{_{semicl}} =
\frac{1}{{\rm x}_{12}^4\ {\rm x}_{34}^{2J}}\  \frac{2 J^2}{N^2} \frac{u^2}{v^2}\,.
\label{4.2}
\ee
This semiclassical result is found in the limit $\sql \gg 1$ 
and $J \sim \sql \gg 1 $. Since it  is given by 
a  product of the three-point chiral primary  correlators 
 divided by their  two-point function each of which  is 
 not \ci{Lee,eden}   renormalized by $\l$ 
 we may conjecture that this expression should  match the large $J$ limit
  of the corresponding gauge
 theory correlator in free $\cal N$=4 SYM  theory. 
 Also, for large  charges  such 4-point correlator
 approaches an extremal one ($J \approx J + 2 + 2 $) 
  and thus,  as   
  was mentioned at the end of section 4.2,   should not be renormalized. 
 
In free $SU(N)$  SYM theory 
the four-point function of  chiral primary gauge theory 
operators ${\cal O}_j (\vec{{\rm x}}, n)$ defined as ($\Phi^k$ are 6 real scalars
and $\n=(n_1,...,n_6)$  is  complex constant vector)
\bea 
&& {\cal O}_j (\vec{{\rm x}}, \n)= \frac{1}{\sqrt{j}} \Big(\frac{8 \pi^2}{\lambda}\Big)^{j/2}
n_{k_1}\dots n_{k_j} {\rm tr} (\Phi^{k_1}\dots \Phi^{k_j})\,,\qquad
{\cal O}_{-j} (\vec{{\rm x}}, \n)= {\cal O}_{j} (\vec{{\rm x}}, \bar \n) \ , 
\label{4.5}\\ 
%
%
&&\sum^6_{k=1} n_k n_k=0\,,\qquad \qquad \sum^6_{k=1} n_k \bar n_k=1\,,
\label{4.6}
\eea
was computed in~\cite{Rayson}
for the  choice of operators with  charges 
 $2, -2, J, -J$
(see also~\cite{Linda1}).
  For  $J\geq4$ it is given by
(we consider  the planar limit $N \gg 1$) 
%
\bea
&&
\langle
{\cal O}_{2}(\vec{{\rm x}}_1, \n_1){\cal O}_{-2} (\vec{{\rm x}}_2, \n_2) 
{\cal O}_{J} (\vec{{\rm x}}_3, \n_3) {\cal O}_{-J}(\vec{{\rm x}}_4, \n_4)\rangle_{_{free\ gauge\
th.}}
\nonumber \\
&& \qquad  =\ 
\Big( \frac{ \n_1 \cdot \n_2}{{\rm x}_{12}^2}\Big)^2 
\Big( \frac{ \n_3 \cdot \n_4}{{\rm x}_{34}^2}\Big)^J
\ {\cal G}^{2,2,J,J} (u, v; p, q) 
\label{4.3}
\eea
Here $u$ and $v$ are again  the conformal cross-ratios~\eqref{1.16} and the function
${\cal G}^{2,2,J,J} (u, v; p, q)$ is given by
\be 
{\cal G}^{2,2,J,J} (u, v; p, q)=1+
\frac{2 J}{N^2} \Big[ p u+ q \frac{u}{v}+ (J-1)\Big(  p q \frac{u^2}{v}
+ p^2 u^2+ q^2 \frac{u^2}{v^2}\Big)\Big]\,, 
\label{4.4}
\ee
where  $p$ and $q$ are defined as 
\be
p =\frac{ (\n_1 \cdot \n_3)(\n_2 \cdot \n_4)}{(\n_1 \cdot \n_2)(\n_3 \cdot \n_4)}\,, \qquad\qquad
q =\frac{ (\n_1 \cdot \n_4)(\n_2 \cdot \n_3)}{(\n_1 \cdot \n_2)(\n_3 \cdot \n_4)}\,.
\label{4.7}
\ee
The term  1 in \rf{4.4}  is  the 
 disconnected   contribution that we  ignored in the previous discussion,
  i.e.
  $\langle
{\cal O}_{2}(\vec{{\rm x}}_1, \n_1){\cal O}_{-2} (\vec{{\rm x}}_2, \n_2) \rangle\langle
{\cal O}_{J} (\vec{{\rm x}}_3, \n_3) {\cal O}_{-J}(\vec{{\rm x}}_4, \n_4)\rangle$; 
it  is useful to keep it here to indicate that the two-point function is assumed 
to be unit-normalized.

Let us specify these expressions to  the operators of interest
\bea
&& {\cal O}_2 (\vec{{\rm x}}_1)=\frac{1}{\sqrt{2}} \frac{4 \pi^2}{\lambda}\ 
{\rm tr}\ Z^2\,, \qquad 
 {\cal O}_{-2} (\vec{{\rm x}}_2)=\frac{1}{\sqrt{2}}  \frac{4 \pi^2}{\lambda}\ 
{\rm tr}\ \bar Z^2\,, \ \ \ \ \ \ \ \ \   Z  \equiv  \Phi^1 + i \Phi^2\ ,   \la{zez}\\
&& {\cal O}_J (\vec{{\rm x}}_3)=\frac{1}{\sqrt{J}} \Big(\frac{4 \pi^2}{\lambda}\Big)^{J/ 2}
{\rm tr}\ Z^J\,, \qquad 
 {\cal O}_{-J} (\vec{{\rm x}}_4)=\frac{1}{\sqrt{J}} \Big(\frac{4 \pi^2}{\lambda}\Big)^{J/ 2}
{\rm tr}\ \bar Z^J\,,
\label{4.7.1}
\eea
%
corresponding to the following choice of $\n_r$
\bea
&&
\n_1= \n_3 =\frac{1}{\sqrt{2}}(1, i, 0, 0, 0, 0)\,,\qquad \qquad 
\n_2= \n_4 =\frac{1}{\sqrt{2}}(1, -i, 0, 0, 0, 0)\,.
\label{4.8}
\eea
Then \rf{4.3} simplifies to 
\be
\langle
{\cal O}_{2}(\vec{{\rm x}}_1){\cal O}_{-2} (\vec{{\rm x}}_2) 
{\cal O}_{J} (\vec{{\rm x}}_3) {\cal O}_{-J}(\vec{{\rm x}}_4)\rangle_{_{free\ gauge\ th.}} 
= \frac{1}{{\rm x}_{12}^4\ {\rm x}_{34}^{2 J}}
\Big[ 1+ \frac{2J}{N^2} \frac{u}{v} +\frac{2 J (J-1)}{N^2} \frac{u^2}{v^2}\Big]\,.
\label{4.9}
\ee
If we now take $J$ to be large the dominant contribution 
to connected $O({1 \ov N^2})$ part of the correlator 
comes from the last term in \rf{4.9}, i.e. 
 we  match    precisely  our 
semiclassical prediction at strong coupling~\eqref{4.2}.

Next, let us attempt to compare~\eqref{4.2}  with  the
 supergravity  result 
 which  should also correspond  to  strong coupling limit of 
 planar gauge theory  correlator  with chiral primary operators  represented 
 by the  appropriate \ci{Lee,af} supergravity scalar modes. 
 Here one finds ~\cite{Linda1} 
\bea
&&\langle
{\cal O}_{2}(\vec{{\rm x}}_1){\cal O}_{-2} (\vec{{\rm x}}_2) 
{\cal O}_{J} (\vec{{\rm x}}_3) {\cal O}_{-J}(\vec{{\rm x}}_4)\rangle_{_{supergr}} 
 = \ \frac{1}{{\rm x}_{12}^4\ {\rm x}_{34}^{2 J}}
\Big[ 1+ \frac{2J}{N^2} \frac{u}{v}  - \frac{1}{N^2} \hD(u,v; J)  \Big] \  \la{4.10}\\
 && \qquad \qquad \qquad \qquad\qquad \qquad \hD (u,v; J) = \frac{2J u^J}{(J-2)!}\
  {\bar D}_{J, J+2, 2, 2} (u, v)
\,.
\label{4.10.1}
\eea
Here the function $\bar D$ is  defined as follows
in terms of the standard four scalar bulk-to-boundary propagator integral in $AdS_5$ 
(with dimensions of  operators being $\Delta_1, ...,
 \Delta_4$)\foot{The expression 
  in
\rf{4.10.1} is a result  of a non-trivial summation of many contributions
 which is the reason why 
 $\bar D$  has  a somewhat ``counter-intuitive''  assignement of its labels compared 
 to  order of the operators in the l.h.s. in \rf{4.10}.}
\bea
&& {\bar D}_{\Delta_1, \Delta_2,\Delta_3,\Delta_4}(u,v) =c_{\Gamma}\ 
\frac{{\rm x}_{13}^{2 \Sigma -2 \Delta_4} \ {\rm x}_{24}^{2 \Delta_2}}
{ {\rm x}_{14}^{2 \Sigma -2 \Delta_1-2 \Delta_4} \ {\rm x}_{34}^{2 
\Sigma-2 \Delta_3-2 \Delta_4}}\ 
D_{\Delta_1, \Delta_2,\Delta_3,\Delta_4}\,,
\label{4.11}\\ 
&& D_{\Delta_1, \Delta_2,\Delta_3,\Delta_4}(\vec{{\rm x}}_1,
 \vec{{\rm x}}_2, \vec{{\rm x}}_3, \vec{{\rm x}}_4)
=\int \frac{d z d^4x}{z^5}\ 
\prod_{i=1}^4 \left[ \frac{z}{z^2+ (\vec{x}-\vec{{\rm x}}_i)^2}\right]^{\Delta_i}\,,
\la{de}\\
&& 
\Sigma\equiv  \ha  \sum ^4_{i=1}\Delta_i \,, \qquad\qquad c_{\Gamma}=\frac{2}{\pi^2} 
\frac{ \Gamma(\Delta_1)\Gamma(\Delta_2)\Gamma(\Delta_3)\Gamma(\Delta_4)}
{\Gamma(\Sigma-2)}\,.
\label{4.12}
\eea
The first two terms in \eqref{4.10} are 
the same as in the free gauge theory expression 
 \rf{4.9}  but instead of the third term in \rf{4.9} we have a complicated 
 ${\hD}$-term.
As we will show 
in  Appendix B, the ${\hD}$-term is {\it subleading}  
for large $J$, i.e. for large $J$ the supergravity 
expression  is dominated by the second  ($\frac{2J}{N^2} \frac{u}{v}$) term. This is  
  in  an apparent contradiction with our 
semiclassical  result  \rf{4.2}.

At the same time, the correspondence 
  between the  supergravity computation  and the string semiclassical computation 
for ``massless''  string modes (e.g., the 
  supergravity chiral primary  scalars)
should be expected on general grounds. Indeed,  
 the supergravity computation in terms 
of scalar particle propagators should admit a 
reformulation in terms 
of  ``first-quantized'' superparticle  path integral  
and 1-d  vertex operators, as it should be an 
 appropriate $\a' \sim  { 1 \ov \sql} \to 0$ limit of the full 
 string computation.\foot{An apparent   difference is 
 that in  the semiclassical string theory discussion we took
 $\sql \gg 1$ with ${J\ov \sql}$=fixed  while  in the supergravity
 analysis  $J$  can take any integer values.  However,   here we are dealing only
 with massless string  modes  (or BPS  states)  and  in both cases 
 $\sql$ is taken to infinity, and also  we are after the leading (positive-power)
 $J^2\gg 1 $ contribution, 
  there is no reason to doubt that the
  supergravity expression  should be valid for all values  of $J$, including 
  $J\sim \sql \to \infty$. The above  arguments  suggesting 
  non-renormalization of this
  particular  correlator  support  this expectation.}
 This  agreement   would be restored if 
 the supergravity expression in \rf{4.10} found in~\cite{Linda1}
is, in fact, missing a   term 
 equivalent to the last term in the free
 gauge theory result \rf{4.9}, i.e.  
\be
\Big(  \langle {\cal O}_{2}(\vec{{\rm x}}_1){\cal O}_{-2} (\vec{{\rm x}}_2) 
{\cal O}_{J} (\vec{{\rm x}}_3) {\cal O}_{-J}(\vec{{\rm x}}_4)\rangle \Big)_{_{supergr. \ extra}}
= \frac{1}{{\rm x}_{12}^4\  {\rm x}_{34}^{2 J}}\frac{2 J (J-1)}{N^2} \frac{u^2}{v^2}
\label{4.13} \ . 
\ee
This term  will 
then    dominate at large $J$ and match the semiclassical 
expression in \rf{4.2}.\footnote{It would  be important to recheck the computation of 
\cite{Linda1} since   comparison 
of our result  with the supergravity one  is a priori  non-trivial. 
In the  computation based directly on 
the  supergravity action, one has to sum, even in the limit of large $J$,  
 over 
many Witten's diagrams corresponding to exchanges of BPS states with large charges.
The  string theory computation  based on vertex operators
 non-trivially ``repackages'' the field theory result,  even 
 when applied  
to ``protected''  correlators.}

%



\section{Discussion}

As was discussed above, 
the semiclassical expression  \rf{1.14},\rf{1.15} 
for the 4-point functions 
may be interpreted as describing the  process in which a ``heavy''
classical  
 closed  string emits  two  ``light''  quantum string modes.
This  can be pictured as a cylindrical world surface  
 with    two  ``light''  world lines 
attached  at different points.
The process  where the
 ``macroscopic'' string first
emits 
one ``light'' mode  which then  decays into two other ``light'' modes
The reason  is  that  the latter process
Indeed,  each 3-point function  entering \rf{1.15}  contains  two 
``heavy'' operators and thus scales as $\sql$  while a correlator  
of  3 ``light'' states will be of order 1.\foot{For example, for 3 chiral
 primary scalars with 
charges  $j_i$ one gets $G_3 = {1 \ov N} \sqrt{ j_1 j_2 j_3}$  \ci{Lee}.}

In general, either on the 
string side or on the dual CFT side,   higher-point correlators of
primary conformal operators are, in principle,  determined 
(via the associative OPE or factorization)
by their 2-point and 3-point  correlation functions. 
Considering the correlator in \rf{1.12}  we may 
factorize it, say, in $13 \to 24$ channel,  getting, symbolically 
\be  \langle \V_{H} \V^*_{H} \V_{L1}\V_{L2} \rangle
 = \sum_A  c_A  \langle \V_{H} \V_{L1}  \V^*_{A} \rangle\  \langle \V_{A} \V^*_{H}
 \V_{L2}\rangle \ 
  \label{v}\ . \ee
 In the  semiclassical (large charge, large $\sql$)
  approximation the sum over the intermediate states 
 will   be dominated  by the  contribution  with  $\V_A= \V_H$, 
 i.e. will be given by  the expression in \rf{1.15}. 
 The same  result should  be found also   if we  factorise in the 
 $12 \to 34$   channel, i.e. 
 \be  \langle \V_{H} \V^*_{H} \V_{L1}\V_{L2} \rangle
 = \sum_A  c_A  \langle \V_{H} \V^*_{H}  \V^*_{A} \rangle\  \langle \V_{A} \V_{L1}
 \V_{L2}\rangle \ .  \label{vv} \ee
 To reproduce 
 the semiclassical result \rf{1.15}  from \rf{vv}  one would  need  to   sum
  over 
 an infinite number of   contributions of  intermediate states. 
One particular type of  contributions
 in \rf{vv}  will be the one  with $\V_A$ as a ``light'' state, which, as was
  mentioned above, is
 subleading in the  semiclassical limit.

 The terms in  \rf{vv}
 with $\V_A$ as a ``heavy'' state, i.e.  $\V_A=\V_{H'}$, 
  should also be  subleading for  large $\sql$.
 Indeed, $\langle \V_{H} \V^*_{H} \V_{H'} \rangle$  and
  $\langle \V_{H'} \V_{L1} \V_{L2} \rangle$ 
 are expected  to  scale  exponentially  in $\sql$  
   at large $\sql$.\foot{The
    leading semiclassical contribution is given by  $e^{-I}$ where 
    $I$  is the string action evaluated on the classical  solution.  
   Strictly
 speaking,  it is not clear a priori  if the exponential factor will 
 always  decay  with large $\sql$  as the  value of the string action 
 on the stationary-point (possibly complex) Euclidean solution
 may not be positive. Still, the expectation of decay 
  is supported  by  the explicit example in \ci{gro}
  of 3 ``heavy'' operators 
  (one  non-BPS, representing rigid circular string  with spins $J_1,J_2$
  and two BMN ones with large spins $J'_1$ and $J''_1$)
  that scales as exp$[ - J_1  
   f({ J_2\ov J_1}, { J'_1\ov J_1}, { J''_1\ov J_1}) ]$
   where $f$ is positive.} 
 Hopefully, their explicit  semiclassical expressions  will    be 
 possible to find using  integrability-based method recently 
  developed in \ci{gro}.\foot{The  case  with 3 ``heavy'' operators 
 is conceptually different from the one considered in \ci{z,rt} and
 here. 
If only two operators carry large quantum numbers the leading 
contribution to the 3-point correlation functions comes from evaluating the 
``light'' vertex operator on the appropriate classical  string solution
on 2-cylider.
On the other hand, if we consider a  correlator with 
3  ``heavy'' operators,  the   corresponding 
semiclassical trajectory will  no longer be directly related to a
  known smooth
 spinning string
solution. Instead, it   should be
 describing a semiclassical 
decay of one large  string into two other 
large strings, with the amplitude proportional to
the exponent of the corresponding value of the classical string action, 
i.e.  $ \sim e^{-a \sql }$  (this  factor may   cancel out if all 3 states are
near-BMN).
This exponential contribution  will be  multiplied also  by 
     other  factors  (not depending on large charges)  in the ``heavy'' 
     vertex operators  
     evaluated on the classical solution. 
    The resulting ``pre-exponential''  factor will scale at least as 
     $ \sql$  so will still  dominate over 
     quantum string corrections, 
      %
 coming, e.g.,  from  the one-loop
fluctuation  determinant, which will be of order 1. 
}

One may also consider  4-point functions involving  more than 2
 ``heavy'' operators. One might 
 expect that 
 $ \langle  \V_{H1} \V_{H2} \V_{H3} \V_L \rangle$
will again scale exponentially with large $\sql$.
At the  same  time, the  factorization relations similar to \rf{v},\rf{vv}   
 seem to  suggest that  at least for $\rx_1 \to \rx_2$, \ $\rx_3 \to \rx_4$ 
(and  particular choice of large charges) 
 the 4-point correlator 
 
 \noindent 
   $ \langle  \V_{H1}(\rx_1)  \V_{H2}(\rx_2) \V_{H3}(\rx_3) \V_{H4}(\rx_4) \rangle$
may   be dominated 
 by the same type of  semiclassical  contribution
  as  in the case of $\langle  \V_{H1} \V_{H2} \V_{L1} \V_{L2} \rangle$, i.e.
by a sum of  products of the semiclassical  3-point functions
with intermediate ``light'' states, i.e.  
$ \langle  \V_{H1} \V_{H2} \V_L \rangle \langle 
\V_L  \V_{H3} \V_{H4} \rangle$.

 As was  already mentioned in \ci{rt} and above, a similar semiclassical 
approach  as  discussed  here   on the example  of the 
4-point functions  can  be  applied  also to  the study of  strong-coupling 
limit of higher-point correlation  functions  containing 
exactly two 
 operators with  large charges. The resulting expressions are  given by 
 direct 
 generalization of \rf{1.15}. It may be of interest to 
 apply them, e.g.,  to the processes  of emission of soft ``gravitons'' 
 by a semiclassical spinning string.
 One  may    also  consider a 
 sum of  such  correlators with the same type of ``light'' state
  corresponding to the exponentiation of the ``light'' vertex operator. 
  This  may be of interest  for understanding the effect of ``back reaction'' of the 
  ``light'' states on the ``heavy'' one.

\section*{Acknowledgments}

We are  grateful to 
  R. Roiban, J. Russo,  L. Uruchurtu  and K. Zarembo for   very   useful discussions
  and suggestions.
  We also thank   N. Drukker,  N. Gromov   and T. McLoughlin    for 
  helpful comments and questions. 
The work of E.I.B. is supported by an STFC fellowship. 

\appendix
\section{Structure of chiral primary  scalar  vertex operator}

Let us review the structure of the full  chiral primary  scalar  vertex operator
as given in \ci{z}  by following \ci{ber} and 
adapting  the  results of  \ci{pvn,Lee}. 
%
Assuming  that the operator is inserted at  point $\vec{\rx}=0$  at the boundary of $AdS_5$ 
we have $({\mathbb X}^{\mathbb M}= (y^m, X^k)$)
\bea
&& \V_j(0) = 
 \cc_j \int d^2 \s \    h_{{\mathbb M} {\mathbb N}}({\mathbb
X})   \ \partial_{\alpha}
{\mathbb X}^{\mathbb M} \partial_{\alpha}{\mathbb X}^{\mathbb N} \
 \label{A1} \\
&&
 = \  \cc_j 
\int d^2 \s\  \Big[  \Big( a\ g_{mn}(y)\ +\  b\  \cD_m \cD_n \Big) \phi_j (y,X)\   \del_\a y^m \del_\a y^n   
   -  \phi_j (y,X)\ \del_\a X_k  \del_\a X_k \Big]
\no\\ 
&& \phi_j(y,X) =  K_j (y) \ \rX^j \ , \ \ \ \ \ \ \ \
K_j = \Big(\frac{z}{z^2 + \vec{x}^2}\Big)^j , \ \ \ \ \ \   \ \ \ \  \rX \equiv X_1 + i X_2 \ , 
\label{3.4.1}\\
&& a = { j-1 \ov j+1} \ , \ \ \ \ \ \   b=  - { 2\ov  j (j+1)}  
\ , \ \ \ \ \
 \cc_j= { \sql  \ov N} { (j+1) \sqrt{j} \ov 8 \pi}  \ . \la{coc}
\eea
Here $y^m=( z,\vec{x} )$ with 
 $ \vec{x}=\{ x^\mu\}  \ (\mu= 0,1,2,3)$  are  $AdS_5$ Poincar\'e patch   coordinates
 with $g_{mn} dy^m dy^n = z^{-2} ( dz^2 + dx^\m dx^\m ) $ and $\cD_m$ are the
  corresponding covariant
 derivatives. 
 $X_k$ are embedding coordinates of $S^5$. 
 $\phi_j$   solves the free 
  scalar  field equation in $AdS_5$ with mass $M^2= j (j-4)$
  with the boundary condition $\phi_j (z,\vec{x})_{z\to 0} \sim \delta^{(4)} (\vec{x}) $.
   The above  expression 
   for $\V_j(0)$ is manifestly $SO(2,4) \times SO(6)$ covariant. 

The  covariant derivatives $\cD_m \cD_n$  act  only on  the 
$K_j(y)$  part of $\phi_j$  and we find 
\bea
&& {\cal D}_{\mu} {\cal D}_{\nu} K_j=
(\partial_{\mu} \partial_{\nu} -\frac{1}{z}\delta_{\mu \nu} \partial_z)K_j=
\Big[-\frac{j }{z^2}\delta_{\mu \nu} +
\frac{4 j (j+1)}{(z^2 +\vec{x}^2)^2}x_{\mu} x_{\nu} \Big] \ K_j\,,
\nonumber\\
&& {\cal D}_{\mu} {\cal D}_z K_j=
(\partial_{\mu} \partial_{z} +\frac{1}{z} \partial_{\mu})K_j=
\frac{2 j (j+1) (z^2 -\vec{x}^2)}{(z^2 +\vec{x}^2)^2}\frac{x_{\mu}}{z}\ K_j\,,
\nonumber\\
&&{\cal D}_z {\cal D}_z K_j =
(\partial_{z} \partial_{z} +\frac{1}{z} \partial_{z})K_j =\Big[
\frac{j^2 }{z^2} -\frac{4 j (j+1) \vec{x}}{(z^2 +\vec{x}^2)^2}\Big] \ K_j\,,
\label{A6}
\eea
where we  split the index $m$ into the $z$-component and the boundary $\mu$-components.
Substituting \eqref{A6} into~\eqref{A1} we  end up with 
the expression quoted in \rf{3.4}, \rf{3.14}, \rf{3.15}, i.e. 
\bea
&&\V_j(0)=\cc_j \int d^2 \s\  K_j \  \rX^j\  U\,,\ \ \ \ \ \ \ \ \ \ \  U= U_1 + U_2 + U_3  \ , 
\label{A7}\\
&&U_1 =   {1 \ov z^2} \big( \del_\a \vec{x} \cdot \del_\a \vec{x} -\del_\a  z \del_\a  z  \big) - \del_\a
X_k \del_\a X_k  \ ,\label{u1} \\ 
&&U_2+U_3 =  \frac{8}{(z^2+\vec{x}^2)^2}
\Big( \big[ \vec{x}^2 (\partial_{\alpha}z)^2 - (\vec{x} \cdot \partial_{\alpha}\vec{x})^2\big]
 + \frac{1}{z} (\vec{x}^2-z^2)(\vec{x} \cdot \partial_{\alpha}\vec{x})
\partial_{\alpha} z\Big)  \  . \la{uu}
\eea
Let us 
 now  evaluate \eqref{A7} on the solution~\eqref{3.3} corresponding to the insertion 
of two ``heavy''  chiral primary operators.  If we choose  $a_1=-a_2$
in \eqref{3.3} we find that 
\be 
U_2 =-U_3 =8  \kappa^2 \frac{\tanh^2 \kappa \tau_e}{\cosh^4 \kappa \tau_e}\big(1-\sinh^2 \kappa \tau_e
\big)\ , \ \ \ \ \ \ \ \ U_2+U_3=0\ , 
\label{A8}
\ee
i.e. the result of using full the vertex operator \rf{A7} in \rf{3.2} is the
 same as using the 
``truncated'' 
operator with $U=U_1$
as in \ci{rt}.
For generic  $a_1$ and $a_2$  we get 
\bea
&& K_j= \frac{a_{12}^j}{
(a_1^2 e^{\kappa \tau_e}+a_{2}^2
 e^{-\kappa \tau_e})^j
 }\,,\ \ \
 \qquad
\rX^j = e^{j \kappa \tau_e}\,, \ \ \ \ \ \ \ \ \ 
U_1 =\frac{2  \kappa^2}{\cosh^2 (\kappa \tau_e)}\,, \nonumber\\
&& U_2 =-2\kappa^2 a_{12}^2 
\Big( \frac{a_1 e^{\kappa \tau_e}+a_2 e^{-\kappa \tau_e}}
{a_1^2 e^{\kappa \tau_e}+a_2^2 e^{-\kappa \tau_e}}
\Big)^2\  \frac{1- \sinh^2 \kappa \tau_e }{\cosh^4  \kappa \tau_e }\,,
\label{A9}\\
&& U_3=2 \kappa^2  a_{12} 
(a_1 e^{\kappa \tau_e}+a_2 e^{-\kappa \tau_e})
        \ \frac{ (a_1 - a_2)^2   - 
 (a_1 e^{\kappa \tau_e}+a_2 e^{-\kappa \tau_e})^2   }
 {(a_1^2 e^{\kappa \tau_e}+a_2^2 e^{-\kappa \tau_e})^2 }
 \ \frac{\sinh \kappa \tau_e}{\cosh^4 \kappa \tau_e}\ .
\nonumber
\eea
This then leads to \rf{3.16}, \rf{3.17}.

Let us now 
evaluate the vertex operator~\eqref{A7} on the solution~\eqref{3.3.4} corresponding 
to the insertion of two  large spin $S$  operators. 
In this case we find 
\bea
&& K_j= \frac{a_{12}^j}{\cosh^j \mu \s\  \big(a_1^2
 e^{\kappa \tau_e}+a_{2}^2 e^{-\kappa \tau_e}\big)^j}\,, \qquad\qquad
X_1+i X_2 =1 \,, \nonumber\\
&& U_1 =\frac{2  \kappa^2}{\cosh^2 \kappa \tau_e}- 2 \mu^2 \tanh^2 \mu \s\,.
\label{A14}
\eea
The expressions for $U_2$ and $U_3$ are rather  complicated, but 
their sum simplifies to 
\be
U_2 +U_3 =-\frac{2  \kappa^2}{\cosh^2 \kappa \tau_e}\ 
\frac{(a_1^2-a_2^2) \ (a_{1}^2 e^{2 \kappa \tau_e} -a_{2}^2 e^{-2 \kappa \tau_e})}
{(a_1^2 e^{\kappa \tau_e}+ a_2^2 e^{-\kappa \tau_e})^2}\,,
\label{A15}
\ee
and again vanishes if $a_1=-a_2$.
Substituting~\eqref{A14}, \eqref{A15} into~\eqref{A1} we 
are led to \rf{3.3.5}.


Let is finish  with a remark on an equivalent form of the 
vertex operator in \rf{A1}  evaluated on $y^m, X_k$   that solve the classical 
string equations of motion (as in the examples we discussed above). 
Consider the covariant derivative term in the integrand in \rf{A1}
\be
A\equiv 
\cD_m \cD_n  ( K_j (y)  \rX^j) 
  \del_\a  y^m \del_\a  y^n = \rX^j  \cD_m \cD_n K_j   \del_\a  y^m \del_\a  y^n 
= \rX^j   \nabla_\a  (\cD_n  K_j)    \del_\a  y^n \ , \la{k}
\ee
where $\nabla_\a \equiv \del_\a  y^n \cD_n $.
Equivalently, 
\be
A=\nabla_\a (\rX^j  \  \cD_n K_j  \ \del_\a y^n )
-\cD_\a \rX^j \ \cD_n K_j \ \del_\a  y^n   
  -  \rX_j  \cD_n K_j \  \nabla_\a \del_\a y^n\ . \la{kk}
\ee
Here the  last term vanishes if $y^n$  satisfies 
the equations of motion (or it can be removed  by a 2d field redefinition). 
Since $\nabla_\a (\rX^j  \  \cD_n K_j  \ \del_\a y^n )
= \del_\a ( \rX^j \del_\a K_j) $ 
We are then left with  the following equivalent form of \rf{A1} 
\be
\V_j(0) = \cc_j \int d^2 \s \  
\Big[  K_j \rX^j \Big(  a\ g_{mn}(y)  \del_\a y^m \del_\a y^n 
- \del_\a X_k  \del_\a X_k  \Big)  + b  \rX^j \del_\a \del_\a  K_j 
\Big] \ ,  
\ee
where  the last term can be written also  as (dropping total derivative)\foot{Note
that the presence of this term that ``mixes'' $AdS_5$ and $S^5$ coordinates in the
``graviton''  vertex operator  may  be  attributed to the mixing with 
 the RR field  fluctuations, requiring some  graviton field redefinitions.} 
\be 
-  b \del_\a \rX^j  \del_\a  K_j= - j  b\rX^{j-1} \del_\a \rX \
 \del_n K_j \ \del_\a y^n    \ . \ee
These expressions are easy to evaluate on given classical solutions.

\section{Large $J$ limit of  ${\hD}$-term in the  supergravity\\ 
expression 
 }
Here we shall study the large $J$ limit of the last  term in eq. ~\eqref{4.10}, i.e. 
\be 
\hD (u, v;J)= \frac{2 Ju^J}{(J-2)!} {\bar D}_{J, J+2,  2, 2}(u, v)\,.
\label{B1}
\ee
We shall   find that it scales as $J^{1/2}$, i.e. is subleading compared to
the second term in \rf{4.10}.

Using eqs.~\eqref{4.11}, \eqref{4.12} and~\eqref{1.16} we get
\be
\hD=\frac{4J(J-1)(J+1)}{\pi^2 } \frac{{\rm x}_{13}^2 
{\rm x}_{24}^4 {\rm x}_{34}^2}{{\rm x}_{14}^2}
{\rm x}_{12}^{2J}\ D_{J, J+2, 2, 2}\,.
\label{B2}
\ee
Since we are interested only in the leading behavior for large $J$ the 
$J$-independent polynomial factors are irrelevant and we can write
\be
\hD\  \sim\  \frac{2J^3}{\pi^2}\  {\rm x}_{12}^{2J} \ D_{J,J,2,2}\,,
\label{B21}
\ee
%
%
where 
\be
D_{J,J,2,2}=\int \frac{d z d^4 x}{z^5}
\Big[ \frac{z}{z^2+ (\vec{x}-\vec{{\rm x}}_1)^2}\ \  \frac{z}{z^2+ (\vec{x}-\vec{{\rm x}}_2)^2} \Big]^J 
\Big[ \frac{z}{z^2+ (\vec{x}-\vec{{\rm x}}_3)^2}\ \ \frac{z}{z^2+ (\vec{x}-\vec{{\rm x}}_4)^2} \Big]^2 
\label{B3}
\ee
 For large $J$  this integral can be evaluated at the  stationary  point
 which  should 
 solve the equations of motion following  from the  ``action''
 (we represent the leading term in the integrand as $e^{- J \ws}$) 
\be
\ws = -  \ln \frac{z}{z^2+ (\vec{x}-\vec{{\rm x}}_1)^2} -
\ln \frac{z}{z^2+ (\vec{x}-\vec{{\rm x}}_2)^2}  \,.
\label{B4}
\ee
We may assume that all the points $\vec{x}_1$ and $\vec{x}_2$ 
in~\eqref{B3} are along the $x_{0e}$-axis. As before we denote their zeroth 
components $a_1$ and $a_2$ and assume $a_1> a_2$.
Then it is straightforward to show that the solution is 
\be
z^2 =(a_1-x_{0e})(x_{0e}-a_2)\,,\ \ \ \ \  \qquad x_1=x_2=x_3=0\,.
\label{B5}
\ee
This solution can be  parametrized as ($a_{12}= a_1-a_2$) 
\bea
 z=\frac{a_{12}}{2 \cosh \tau}\,,\qquad 
  x_{0e}=\frac{a_{12}}{2} \tanh \tau +\frac{1}{2}(a_1+a_2)\,,
   \qquad  x_1=x_2=x_3=0\,.
\label{B6}
\eea
Note that this is precisely the point-particle 
solution in \eqref{3.3} (with $\k=1$).  
Since this stationary  point   solution is not an isolated  point but 
a line 
parametrized by  $\tau$,   the integral over the $AdS_5$ space in \rf{B3}
should be reduced to the integral over $\tau$. 
This is done by relating the integration 
measure to the induced meausure on the curve
\be
\int \frac{d z d^4 x}{z^5} =\int d \tau \sqrt{g_{ind}} =\int d \tau \,, 
\label{B7}
\ee
where  we used that, as follows  from  \eqref{B6}, the determinant
 of the induced metric on the curve
 $g_{ind}$ is equal to 1. 
On the stationary  point solution the ``action'' in \rf{B4} is 
$
\ws=2 \ln a_{12} = 2\ln {\rm x}_{12}$
so that we obtain
\be
D_{J,J,2,2} \ \sim \ \frac{1}{{\rm x}_{12}^{2J}} \int_{-\infty}^{\infty} d \tau
\Big[ \frac{z}{z^2+ (\vec{x}-\vec{{\rm x}}_3)^2} \ \
 \frac{z}{z^2+ (\vec{x}-\vec{{\rm x}}_4)^2} \Big]^2 \  Q\,,
\label{B10}
\ee
where $Q$ is the ``one-loop'' determinant over the fluctuations
around the solution \eqref{B6}
\be
Q =\int d^5 \delta y\ \ 
e^{-\ha J \del_m \del_n \ws \  \delta y^m  \delta y^n }\  \sim \ 
{J^{-5/2}}\  {\rm det}^{-1/2}  (\del_m \del_n \ws)  \ , 
\label{B11}
\ee
where  $\delta y^n= (\delta z, \delta \vec{x})$. 
%
%
The integrand in \rf{B10} has to be evaluated on the solution~\eqref{B6}. 
The $\tau$-integral is $J$-independent, and thus 
the  leading $J$  dependence of \rf{B10}   is given by 
%
\be 
D_{J,J,2,2}\  \sim\  \frac{J^{-5/2} }{ {\rm x}_{12}^{2J}}\,.
\label{B13}
\ee
Substituting this  into~\eqref{B21} gives
\be
\hD   \sim \  {J^{1/2}}\,.
\label{B14}
\ee
This shows that $\hD$ is subleading compared  to~\eqref{4.13}.



\end{document}